\documentclass[preprint,12pt]{elsarticle}

\usepackage[pdfborder={0 0 0}]{hyperref}
\usepackage[margin=1in]{geometry}
\usepackage{lineno}
\usepackage{graphicx}
\usepackage{amsfonts}
\usepackage{epstopdf}
\usepackage{mathtools}
\usepackage{array}
\usepackage{latexsym}
\usepackage{enumitem}
\usepackage[utf8]{inputenc}
\usepackage{float}
\usepackage{hyperref}
\usepackage{subcaption}
\usepackage{algorithm}
\usepackage[noend]{algpseudocode}
\usepackage{tabularx}
\usepackage{bm}
\usepackage{mathrsfs}
\usepackage{multirow}

\hypersetup{
  colorlinks   = true, %
  urlcolor     = blue, %
  linkcolor    = blue, %
  citecolor   = red %
}

\usepackage{color}
\usepackage{xspace}
\usepackage{amssymb}

\graphicspath{{figures/}}

\begin{document}
\begin{frontmatter}

\title{CAD-compatible structural shape optimization with a movable B\'{e}zier tetrahedral mesh}

\author[BUW]{Jorge L\'{o}pez}
\author[BUW]{Cosmin Anitescu}
\author[TDTU,TDTUCE]{Timon Rabczuk\corref{cor1}}
\ead{timon.rabczuk@tdtu.edu.vn}

\address[TDTU]{Division of Computational Mechanics, Ton Duc Thang University, Ho Chi Minh City, Viet Nam}
\address[TDTUCE]{Faculty of Civil Engineering, Ton Duc Thang University, Ho Chi Minh City, Viet Nam}
\address[BUW]{Institute of Structural Mechanics, Bauhaus-Universit\"{a}t Weimar, Marienstr. 15, D-99423 Weimar, Germany}

\cortext[cor1]{Corresponding author}

\begin{abstract}
This paper presents the development of a complete CAD-compatible framework for structural shape optimization in 3D. The boundaries of the domain are described using NURBS while the interior is discretized with B\'{e}zier tetrahedra. The tetrahedral mesh is obtained from the mesh generator software Gmsh. A methodology to reconstruct the NURBS surfaces from the triangular faces of the boundary mesh is presented. The description of the boundary is used for the computation of the analytical sensitivities with respect to the control points employed in surface design. Further, the mesh is updated at each iteration of the structural optimization process by a pseudo-elastic moving mesh method. In this procedure, the existing mesh is deformed to match the updated surface and therefore reduces the need for remeshing. Numerical examples are presented to test the performance of the proposed method. The use of the movable mesh technique results in a considerable decrease in the computational effort for the numerical examples.

\end{abstract}


\end{frontmatter}

\section{Introduction}
\label{S:1}

Modeling and analysis have typically been distinguished as two different yet complementary phases for the engineering practice. However, for the analysis using the conventional numerical methods, a discrete mesh is essential for the geometries that have been modeled using Computer-Aided Design (CAD). Various methods exist in the literature \cite{Zienkiewicz2005,Belytschko2001} to discretize the geometry and the solution fields. In this regard, Isogeometric Analysis (IGA) \cite{Hughes2005} emerged as an alternative to Finite Element Method (FEM), proposing the merging of the analysis and design domains. In this framework, the same basis functions which are employed in CAD to describe the geometry are also used for the approximation of the solution fields at the analysis stage.

In the context of shape optimization, IGA provides the opportunity to represent complex geometries with few design variables \cite{Wall2008,Qian2010, Manh2011,Nortoft2013,Park2013,Yoon2013}. This is in contrast to the node-based optimization approach using FEM, where element vertices are used as design variables. Moreover, B-splines and NURBS basis functions, commonly used in CAD software, offer a variety of refinement algorithms that keep the geometry intact.

Even though IGA has several advantages, some difficulties have been identified particularly in case of three-dimensional domains. First, the CAD models might parameterize only the boundaries and not the interior of the domain. A CAD model frequently uses a collection of connected surfaces to represent the geometry, omitting the information required to parametrize the volume. Also, complex geometries cannot be represented with tensor product basis functions. Often, a designer will model geometries that are not analysis-suitable. Furthermore, compatibility between the different CAD software packages and the analysis platforms is crucial.
Considering this, in the context of shape optimization, a CAD-compatible model must meet the following requirements:
\begin{itemize}
    \item It should employ commonly used CAD basis functions to parameterize the boundary of the geometry.
    \item It must be capable of modelling complex geometries.
    \item Exchange of information between the geometry and the analysis software should be possible. 
    \item It should be able to maintain a link between the design variables and the analysis mesh.
\end{itemize}

To overcome the issues discussed above, we broadly present the features of the proposed approach in the context of CAD-compatible shape optimization. First, we propose to use the IGES exchange format to communicate between the design and the analysis mesh. Second, our approach uses B\'{e}zier elements for the discretizetion of the domain, where the geometry can be exactly represented in a CAD-compatible format. This might appear an extension of the NURBS Enhanced Finite Element Method (NEFEM) \cite{Sevilla2008,Sevilla2011a,Sevilla2011b} with the use of IGA. However, in NEFEM the parent element is modified for the curved elements at the boundaries, and therefore the numerical quadrature is altered for the integration of those elements, which is in contrast to the method proposed in this paper. In this work, we provide a linkage between CAD-based design and optimization analysis in a more flexible way than what is provided by standard IGA. The NURBS-based IGA still has some difficulties with volumetric mesh generation since the problem is essentially equivalent to that of hexahedral meshing. Moreover, multiple patches or parameterization singularities are still required for classic IGA. In this regard, the B\'{e}zier tetrahedra can function as a bridge between surface CAD models and
volumetric discretizations.

To make the proposed method efficient for 3D models, we develop a method based on the point inversion algorithm \cite{Piegl2012} to reconstruct the NURBS boundaries. This approach is particularly important in three-dimensional domains, where converting a NURBS surface into triangular elements requires that the degree of the triangle elements is at least twice of the degree of the tensor product description of the initial surface \cite{Goldman1987}, thereby creating a large number of control points. By contrast, the link between the boundary representation and the analysis mesh is more easily achieved for two-dimensional problems by using $C^0$ continuous curve segments as one side of individual triangular elements \cite{Engvall2016,Qian2010,Lopez2019}. 

Lastly, the proposed approach exploits the advantages of linking the NURBS boundary with the B\'{e}zier tetrahedral mesh. The sensitivities can be computed analytically in the analysis mesh with respect to the coordinates of the NURBS control points. Moreover, it provides the possibility to update the mesh without remeshing at each iteration, avoiding changes in the connectivity of the elements and resulting in faster convergence rates without oscillations.

The use of simplex elements in an isogeometric framework has been of interest in recent years. Similar work that has already been carried out include the implementation of  B\'{e}zier triangles in the framework of IGA for applications in vibrations \cite{Liu2018a}, thin plates and shells \cite{Ludwig2019,Zareh2019}, damage mechanics\cite{Liu2018b} and phase separation \cite{Zhang2019}. The increase of continuity to $C^1$ B\'{e}zier triangles has been studied in \cite{Jaxon2014,Xia2015}. Further extensions of 3D implementations using B\'{e}zier tetrahedral meshes have been developed in \cite{Xia2017,Kadapa2019}, and for more general B\'{e}zier elements in \cite{Engvall2017}. In the context of shape optimization, B\'{e}zier triangles have been implemented for the design and optimization of 2D auxetic materials in \cite{Wang2018}, and for stress constraints in \cite{Lopez2019}. Also, the implementation of the projection procedures to approximate the B\'{e}zier elements faces to the NURBS surface has been shown in \cite{Engvall2017}. Lastly, some work on moving mesh techniques have been commonly presented for fluid-structure interaction \cite{Alauzet2014,Alauzet2016,Stein2003,Stein2004,Tezduyar2007}. In the context of optimization, the moving mesh approach has been used in \cite{Belegundu1988,Liu2008,Wang2018}, where the update of the mesh by an elastic body model has been preferred. 

The remainder of the paper is organized as follows. In Section 2, we introduce the basics of Computer Aided Graphic Design (CAGD). We  discuss the basics of elements and basis functions typically used in this modeling approach. In Section 3, an overview of the construction of volumetric objects using the IGES exchange format is given. In Section 4, we describe the approach to reconstruct the NURBS boundaries and the method for smoothing the weights in the interior of the domain. In Section 5, the implementation of the proposed approach for shape optimization is introduced and further explained. The analytical sensitivity analysis and the method to update the mesh based on a pseudo-elastic mesh is also presented. Numerical examples illustrating the performance of the proposed approach are presented in Section 6, followed by concluding remarks and possibilities for future work.

\section{Fundamentals of CAGD}
\label{S:2}
\subsection{Tetrahedral elements}
Unstructured meshes, typically tetrahedral elements for three-dimensional modeling, do not present a regular pattern on the topology or connectivity of their elements. For this reason, the storage of the connectivity between elements is required, which implies higher computational cost in comparison to structured meshes. However, their flexibility to model complex geometries gives them an advantage over their structured counterparts. In this section, we introduce the basic concepts of tetrahedral elements typically used in CAD.

\subsubsection{Barycentric coordinate system}
\begin{figure}[t]
    \centering
    \includegraphics[width=0.6\textwidth]{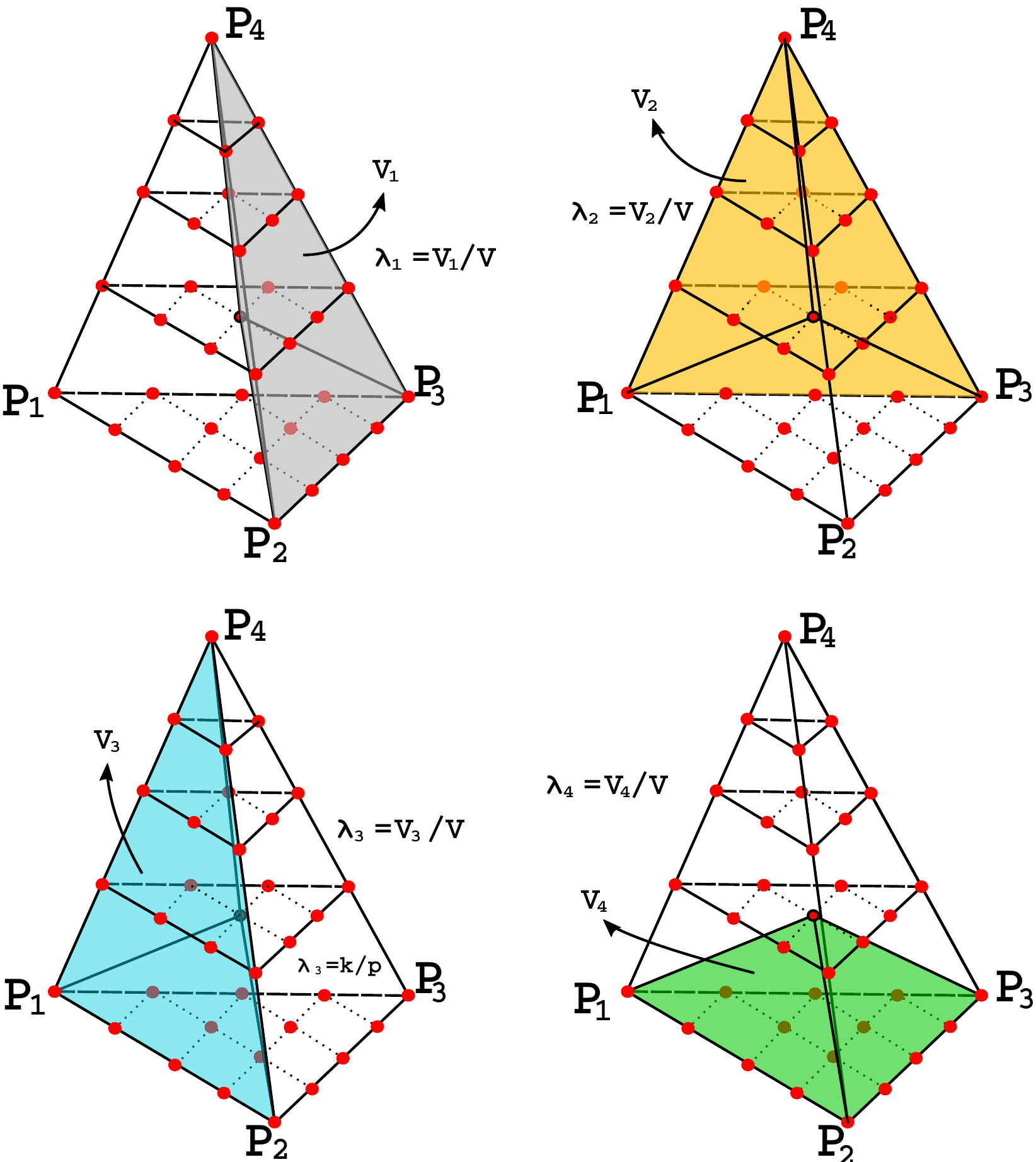}
    \caption{Barycentric coordinate system in a tetrahedron.}
    \label{fig:Tetrahedron_BarCoord}
\end{figure}

First, we describe the barycentric coordinates for a tetrahedron, which later will be used to define basis functions over the element. This system, first introduced by M\"{o}bius \cite{Mobius1827}, determines the location of a point in $\mathbb{R}^3$ as the barycenter of a tetrahedron. Namely, the value of masses located at the 4 vertices of the tetrahedron are given, such that the center of mass of the element can be located. Keeping this interpretation in mind, the barycentric coordinates $\bm{\lambda} = [\lambda_1,\lambda_2,\lambda_3,\lambda_4]$ can be written in terms of the volume of a tetrahedron as:

\begin{equation}
    \lambda_i = \frac{V_i}{V},
\end{equation}
where $V$ denotes the volume of the element and $V_i$ is the volume of the tetrahedron formed by the point of interest and the three opposite vertices to the vertex $i$, as shown in Fig.~\ref{fig:Tetrahedron_BarCoord}. Let us consider a tetrahedron $\bm{T}$ in the Euclidean space $\mathbb{R}^3$ with vertices at $\bm{P}_1$, $\bm{P}_2$, $\bm{P}_3$ and $\bm{P}_4$. Using the barycentric coordinate system, any point $\bm{P}\in\mathbb{R}^3$ can be defined as a linear combination of the vertices of the tetrahedron with the barycentric coordinates $\bm{\lambda}=[\lambda_1,$ $\lambda_2,$ $\lambda_3,$ $\lambda_4]$:

\begin{equation}
    \bm{P}(\bm{\lambda}) = \lambda_1\bm{P}_1 + \lambda_2\bm{P}_2 + \lambda_3\bm{P}_3 + \lambda_4\bm{P}_4,
\end{equation}
satisfying:
\begin{equation}
    \lambda_1 + \lambda_2 + \lambda_3 + \lambda_4 = 1.
\end{equation}

\subsubsection{B\'{e}zier tetrahedra}
\label{S:2.2}
B\'{e}zier curves are parametric curves defined as a linear combination of control points and Bernstein polynomials. Bernstein polynomials can be defined over tetrahedral domains by making use of the previously introduced barycentric coordinates $\bm{\lambda}$. We consider the parent element in the space $(\xi,\eta,\zeta)$ with vertices as shown in Fig.~\ref{fig:Tet_Mapping}. Some of the properties exhibited by the B\'{e}zier basis are partition of unity, pointwise nonnegativity and endpoint interpolation \cite{Lai2007,Engvall2017}. The B\'{e}zier-Bernstein basis functions of degree $p$ over the parent tetrahedron are defined as:

\begin{figure}[t]
    \centering
    \includegraphics[width=0.6\textwidth]{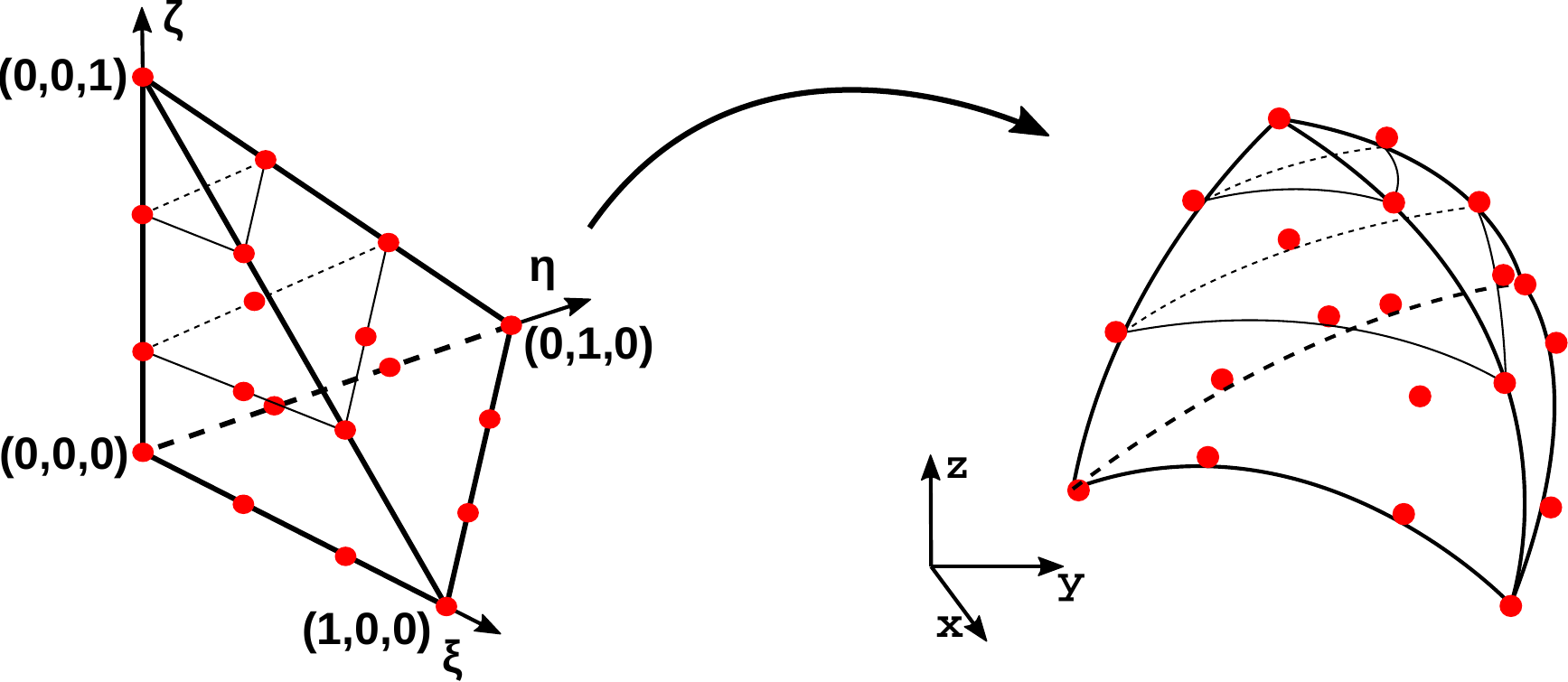}
    \caption{Parent element and its mapping into the physical space.}
    \label{fig:Tet_Mapping}
\end{figure}

\begin{equation}
    B_{ijkl}^p(\bm{\lambda}) = \frac{p!}{i!j!k!l!}\lambda_1^i \lambda_2^j \lambda_3^k \lambda_4^l,
\end{equation}
and its derivatives as:
\begin{equation}
    \begin{array}{r@{}l}
         \frac{d}{d\lambda_1}B_{ijkl}^p(\bm{\lambda}) &= \frac{p!}{(i-1)!j!k!l!}\lambda_1^{i-1} \lambda_2^j \lambda_3^k \lambda_4^l, \vspace{5pt}\\
         \frac{d}{d\lambda_2}B_{ijkl}^p(\bm{\lambda}) &= \frac{p!}{i!(j-1)!k!l!}\lambda_1^i \lambda_2^{j-1} \lambda_3^k \lambda_4^l, \vspace{5pt}\\
         \frac{d}{d\lambda_3}B_{ijkl}^p(\bm{\lambda}) &= \frac{p!}{i!j!(k-1)!l!}\lambda_1^i \lambda_2^j \lambda_3^{k-1} \lambda_4^l, 
         \vspace{5pt}\\
         \frac{d}{d\lambda_4}B_{ijkl}^p(\bm{\lambda}) &= \frac{p!}{i!j!k!(l-1)!}\lambda_1^i \lambda_2^j \lambda_3^k \lambda_4^{l-1},
    \end{array}
\end{equation}
where $\lambda_1$, $\lambda_2$, $\lambda_3$, $\lambda_4$ are the barycentric coordinates on the reference element; and the indexes $i+j+k+l=p$. Usually, the derivatives of the basis functions are computed with respect to the parametric coordinates as:
\begin{equation}\label{eq:basis_ders}
    \begin{array}{r@{}l}
         \frac{d}{d\xi}B_{ijkl}^p(\bm{\lambda}) &=  \frac{\partial\lambda_1}{\partial\xi}\frac{d}{d\lambda_1}B_{ijkl}^p(\bm{\lambda}) + \frac{\partial\lambda_2}{\partial\xi}\frac{d}{d\lambda_2}B_{ijkl}^p(\bm{\lambda}) + \frac{\partial\lambda_3}{\partial\xi}\frac{d}{d\lambda_3}B_{ijkl}^p(\bm{\lambda}) + \frac{\partial\lambda_4}{\partial\xi}\frac{d}{d\lambda_4}B_{ijkl}^p(\bm{\lambda}), \vspace{8pt}\\
         
         \frac{d}{d\eta}B_{ijkl}^p(\bm{\lambda}) &=  \frac{\partial\lambda_1}{\partial\eta}\frac{d}{d\lambda_1}B_{ijkl}^p(\bm{\lambda}) + \frac{\partial\lambda_2}{\partial\eta}\frac{d}{d\lambda_2}B_{ijkl}^p(\bm{\lambda}) + \frac{\partial\lambda_3}{\partial\eta}\frac{d}{d\lambda_3}B_{ijkl}^p(\bm{\lambda}) + \frac{\partial\lambda_4}{\partial\eta}\frac{d}{d\lambda_4}B_{ijkl}^p(\bm{\lambda}), \vspace{8pt}\\
         
         \frac{d}{d\zeta}B_{ijkl}^p(\bm{\lambda}) &=  \frac{\partial\lambda_1}{\partial\zeta}\frac{d}{d\lambda_1}B_{ijkl}^p(\bm{\lambda}) + \frac{\partial\lambda_2}{\partial\zeta}\frac{d}{d\lambda_2}B_{ijkl}^p(\bm{\lambda}) + \frac{\partial\lambda_3}{\partial\zeta}\frac{d}{d\lambda_3}B_{ijkl}^p(\bm{\lambda}) + \frac{\partial\lambda_4}{\partial\zeta}\frac{d}{d\lambda_4}B_{ijkl}^p(\bm{\lambda}),\\
    \end{array}
\end{equation}

For the parent tetrahedra from Fig.~\ref{fig:Tet_Mapping} we use the following relations:
\begin{equation}
\begin{split}
\lambda_1 &= -\xi -\eta -\zeta +1, \\
\lambda_2 &= \xi, \quad \lambda_3 = \eta, \quad \lambda_4 = \zeta.
\end{split}
\end{equation}

Using the vertex coordinates from the parent element, Eq.~\eqref{eq:basis_ders} can be written as:
\begin{equation}
    \begin{array}{r@{}l}
         \frac{d}{d\xi}B_{ijkl}^p(\bm{\lambda}) &=  -\frac{d}{d\lambda_1}B_{ijkl}^p(\bm{\lambda}) + \frac{d}{d\lambda_2}B_{ijkl}^p(\bm{\lambda}), \vspace{8pt}\\
         
         \frac{d}{d\eta}B_{ijkl}^p(\bm{\lambda}) &=  -\frac{d}{d\lambda_1}B_{ijkl}^p(\bm{\lambda}) + \frac{d}{d\lambda_3}B_{ijkl}^p(\bm{\lambda}), \vspace{8pt}\\
         
         \frac{d}{d\zeta}B_{ijkl}^p(\bm{\lambda}) &=  -\frac{d}{d\lambda_1}B_{ijkl}^p(\bm{\lambda}) + \frac{d}{d\lambda_4}B_{ijkl}^p(\bm{\lambda}).\\
    \end{array}
\end{equation}

\begin{figure}[t]
 \hspace{50pt}
 \begin{subfigure}[b] {0.13\textwidth}
    \centering
    \includegraphics[width=\textwidth]{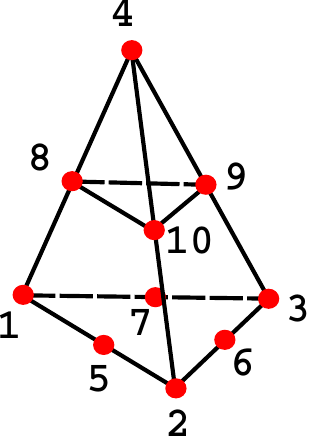}
    \caption{}
    \end{subfigure}\hfill
    \begin{subfigure}[b] {0.2\textwidth}
   \centering
   \includegraphics[width=\textwidth]{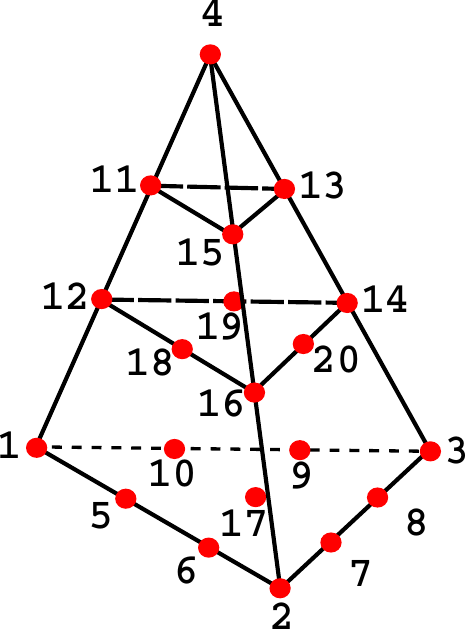}\hfill
   \caption{}
 \end{subfigure}\hfill
 \begin{subfigure}[b] {0.24\textwidth}
   \centering
   \includegraphics[width=\textwidth]{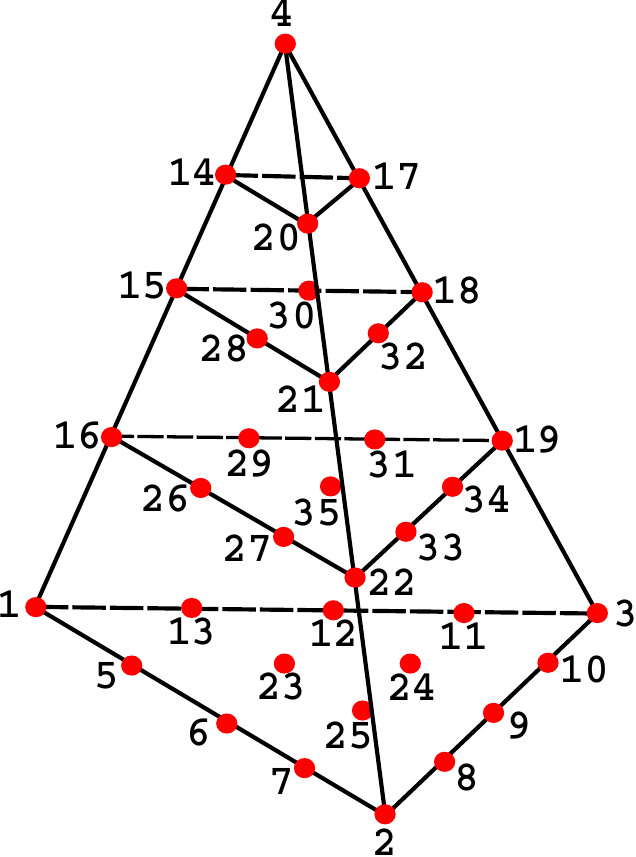}\hfill
   \caption{}
 \end{subfigure}\hspace{50pt}
    \caption{Node numbering of a a) quadratic, b) cubic and c) quartic  B\'{ezier tetrahedron}.}
    \label{fig:CubicTetNumbering}
\end{figure}

Fig.~\ref{fig:CubicTetNumbering} shows the convention used to order the basis functions for a cubic tetrahedral parent element. Following this convention, a tetrahedron in the physical space (see Fig.~\ref{fig:Tet_Mapping}) is described as the linear combination of such shape functions with the control points in the physical space $P_{ijkl}$:

\begin{equation}
    T(\bm{\lambda}) = \sum_{i+j+k+l=p} B_{ijkl}^p(\bm{\lambda}) P_{ijkl}.
\end{equation}

In the case that rational basis functions are used, the tetrahedron in the physical space is defined by:

\begin{equation}
    T(\bm{\lambda}) =  \sum_{i+j+k+l=p} R_{ijkl}^p(\bm{\lambda}) w_{ijkl} P_{ijkl},
\end{equation}
where $w_{ijkl}$ are the weights and the rational basis $R_{ijkl}^p(\bm{\lambda})$ are computed by:
\begin{equation}
    R_{ijkl}^p(\bm{\lambda}) = \frac{B_{ijkl}^p(\bm{\lambda})w_{ijkl}}{W(\bm{\lambda})},
\end{equation}
where $W(\bm{\lambda})$ is the weight function:
\begin{equation}
    W(\bm{\lambda}) = \sum_{i+j+k+l=p} B_{ijkl}^p(\bm{\lambda}) w_{ijkl}.
\end{equation}
 
\subsubsection{Lagrange Tetrahedra}
Though B\'{e}zier tetrahedra are commonly used in the CAD community, mesh generation programs typically generate Lagrange elements, mainly because of the interpolatory property at all nodes \cite{Schillinger2016,Zienkiewicz2005}. Thus, the control points of these elements can be located exactly over the boundaries of CAD models, which usually use NURBS for the geometry description.

\begin{figure}[t]
    \centering
    \includegraphics[width=0.5\textwidth]{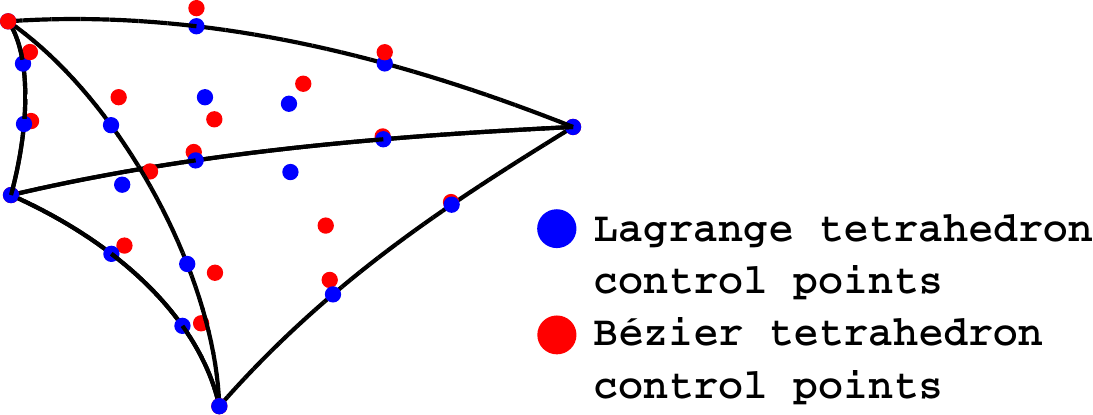}
    \caption{Lagrange and B\'{e}zier control points required to model exactly the same tetrahedron.}
    \label{fig:Tet_Lagrange_Bezier}
\end{figure}

The Lagrange basis of degree $p$ over the same parent element as in the case of B\'{e}zier tetrahedra, are defined as \cite{Zienkiewicz2005}:
\begin{equation}
    L_{ijkl}^p(\lambda_1, \lambda_2, \lambda_3, \lambda_4)=\ell_i(\lambda_1)\ell_j(\lambda_2)\ell_k(\lambda_3)\ell_l(\lambda_4),
\end{equation}
with $i+j+k+l=p$. The functions $\ell_t(\lambda_s)$ are one-dimensional Lagrange polynomials of degree $t$, defined by:
\begin{equation}
    \ell_t(\lambda_s) = \prod_{\substack{r=0 }}^{t-1} \left(\frac{p\lambda_s-r}{t-r}\right).
\end{equation}
The same physical tetrahedron as in the last subsection $\bm{T}(\bm{\lambda})$ is in this case described as a linear combination of the Lagrange basis and the corresponding control points:

\begin{equation}
    T(\bm{\lambda}) = \sum_{i+j+k+l=p} L_{ijkl}^p(\bm{\lambda}) P^{L}_{ijkl},
\end{equation}
where $P^{L}_{ijkl}$ are the control points of the Lagrange tetrahedron. Even though the basis and control points are different, B\'{e}zier and Lagrange basis can be used to model geometrically the same physical tetrahedron, provided that all the weights are equal to one, as shown in Fig.~\ref{fig:Tet_Lagrange_Bezier}.

\subsubsection{Mesh size control}
\begin{figure}[t]
    \centering
    \includegraphics[width=0.6\textwidth]{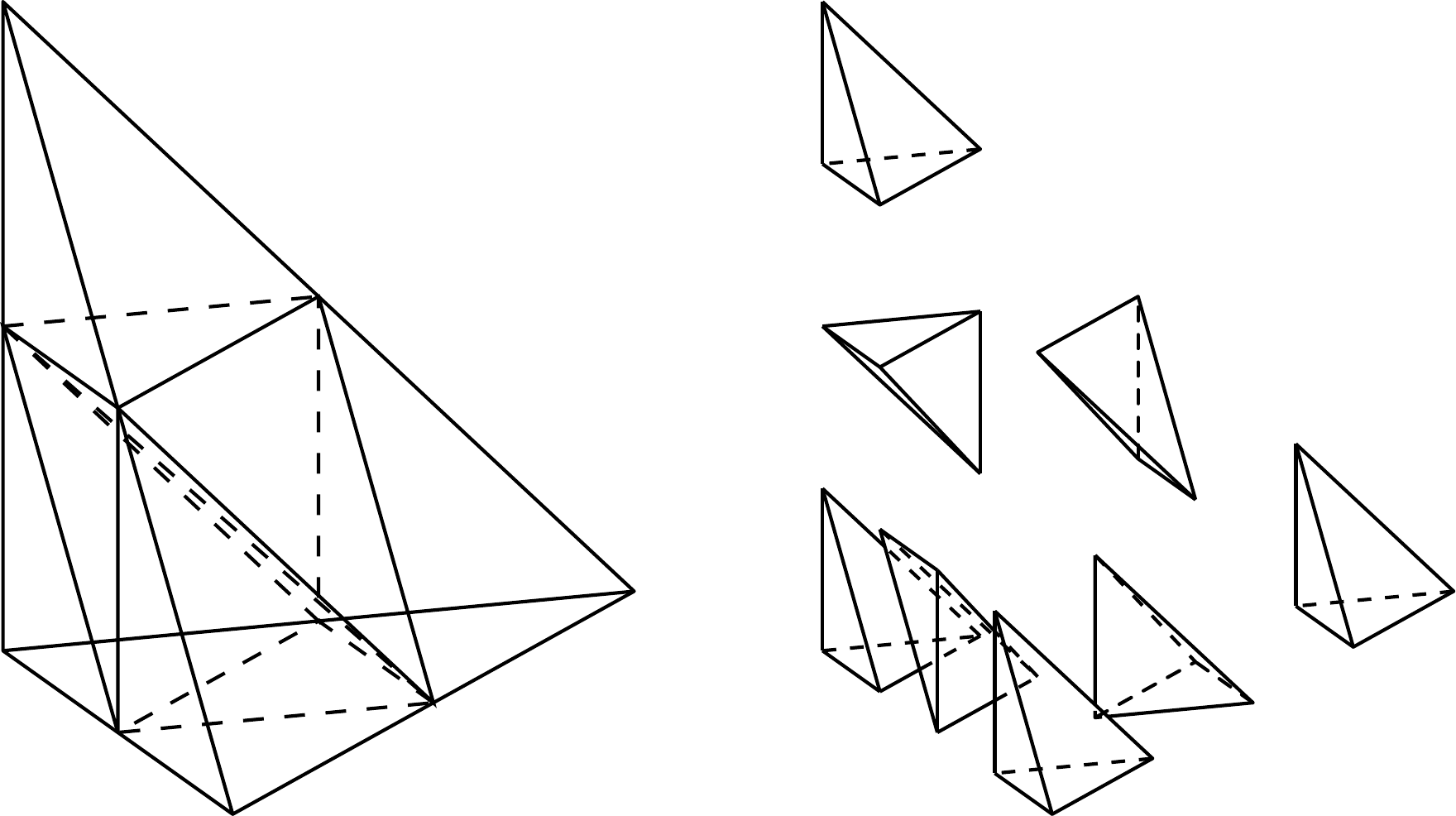}
    \caption{Refinement by splitting of a tetrahedron. One tetrahedron is split into eight tetrahedra.}
    \label{fig:Tet_Refinement}
\end{figure}

In order to control the size of the tetrahedral mesh, two different approaches can be used together or separately. First the mesh size can be controlled by setting a characteristic length $h_l$ at diferent points in the domain.In this case, $h_l$ indicates the radius of a circumscribed sphere for a tetrahedron at some location of the mesh. That is, the size of the mesh is obtained by linearly interpolating the values of all given $h_l$. In this work, a homogeoneous $h_l$ is given for all vertices of the boundary.

Additionally, the mesh can be further refined by splitting each element into smaller tetrahedra. Fig.~\ref{fig:Tet_Refinement} shows the split of an element into eight elements by cutting planes through the middle points of each edge of the tetrahedron.

\subsection{NURBS}
\label{S:2.1}

\begin{figure}[t]
   \centering
   \begin{subfigure}[b] {0.49\textwidth}
   \centering
   \includegraphics[width=\textwidth]{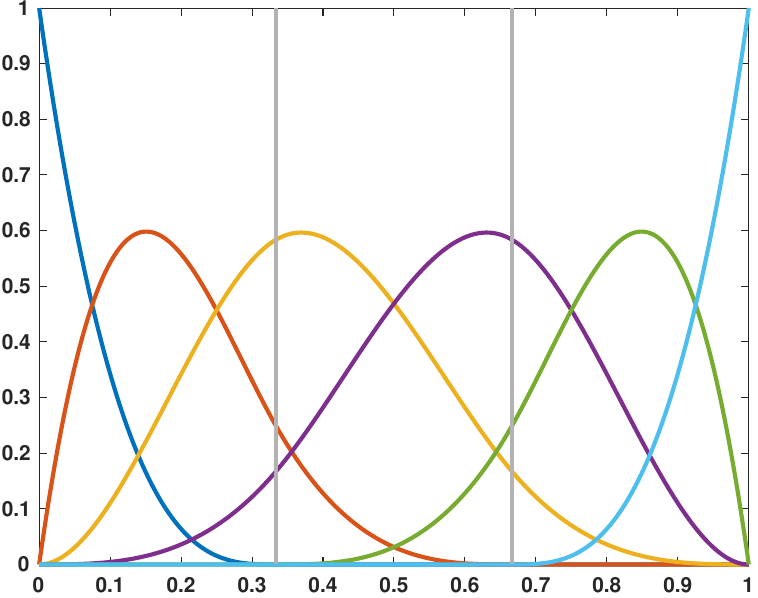}\label{fig:BSpline_Basis1D}\hfill
   \caption{}
 \end{subfigure}\hspace{5pt}
 \begin{subfigure}[b] {0.48\textwidth}
   \centering
   \includegraphics[width=\textwidth]{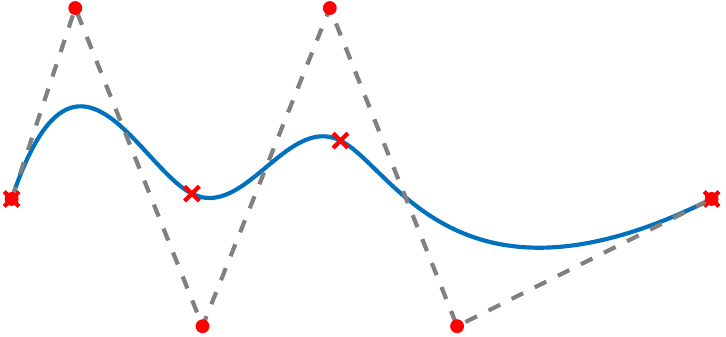}\label{fig:BSpline_1D}\hfill
   \caption{}
 \end{subfigure}\hspace{5pt}
\caption{a) Cubic B-spline basis functions over the parameter space with knot vector $\bm{\xi} = [0,0,0,0,\frac{1}{3},\frac{2}{3},1,1,1,1]$. b) B-Spline curve with its control points, the red crosses show the knot span interfaces in the physical space.}
\label{fig:Basis_Bspline_1D}
\end{figure}

Let $\bm{\Xi} = [\hat{\xi}_1,\hat{\xi}_2, ... , \hat{\xi}_{n}, ...\hat{\xi}_{n+p+1}] $ be a vector of non-decreasing parameter values known as the knot vector, where $n$ is the number of basis functions associated to $\bm{\hat{\xi}}$, and $\hat{p}$ is the degree of those functions. A NURBS curve is a parametric curve defined by:

\begin{equation}
    \bm{C}(\hat{\xi}) = \frac{\sum_{i=1}^{n} N_{i,p}\hat{w}_{i}\hat{P}_{i}}{\sum_{j=1}^{n}N_{j,p}\hat{w}_{j}},
\end{equation}
where $\hat{w}_i$ is the $ith$ weight of the NURBS curve, $\hat{P}_i$ the corresponding NURBS control point and $N_{i,p}$ are B-Spline basis functions of degree $\hat{p}$. B-Splines are piecewise polynomials obtained recursively by the Cox-de-Boor formula:

\begin{equation}
    N_{i,0}(\hat{\xi}) = \begin{cases} 1 \hspace{5pt} \mbox{if} \hspace{5pt} \hat{\xi}_i \leq \hat{\xi} < \hat{\xi}_{i+1}, \\
    0 \hspace{5pt} \mbox{otherwise},
    \end{cases}
\end{equation}

\begin{equation}
    N_{i,p}(\hat{\xi}) = \frac{\hat{\xi} - \hat{\xi}_i}{\hat{\xi}_{i+\hat{p}} - \hat{\xi}_i}N_{i,\hat{p}-1}(\hat{\xi}) + \frac{\hat{\xi}_{i+\hat{p}+1} - \hat{\xi}}{\hat{\xi}_{i+\hat{p}+1} - \hat{\xi}_{i+1}}N_{i+1,\hat{p}-1}(\hat{\xi}).
\end{equation}

Fig.~\ref{fig:Basis_Bspline_1D} shows the cubic B-Splines basis functions over the parameter space $\hat{\xi} \in [0,1]$. A NURBS surface can be obtained by applying tensor product of two NURBS curves with knot vectors $\bm{\Xi} = [\hat{\xi}_1,\hat{\xi}_2, ... , \hat{\xi}_{n}, ...\hat{\xi}_{n+p+1}] $ and $\bm{H} = [\hat{\eta}_1,\hat{\eta}_2, ... , \hat{\eta}_{n}, ...\hat{\eta}_{m+q+1}] $as:

\begin{equation}
    \bm{S}(\hat{\xi},\hat{\eta}) = \sum_{i=1}^{n} \sum_{j=1}^{m} \hat{R}_{i,j}^{\hat{p},\hat{q}}(\hat{\xi},\hat{\eta})\hat{w}_{i,j}\hat{P}_{i,j},
\end{equation}

\begin{figure}[t]
   \centering
   \begin{subfigure}[b] {0.5\textwidth}
   \centering
   \includegraphics[width=\textwidth]{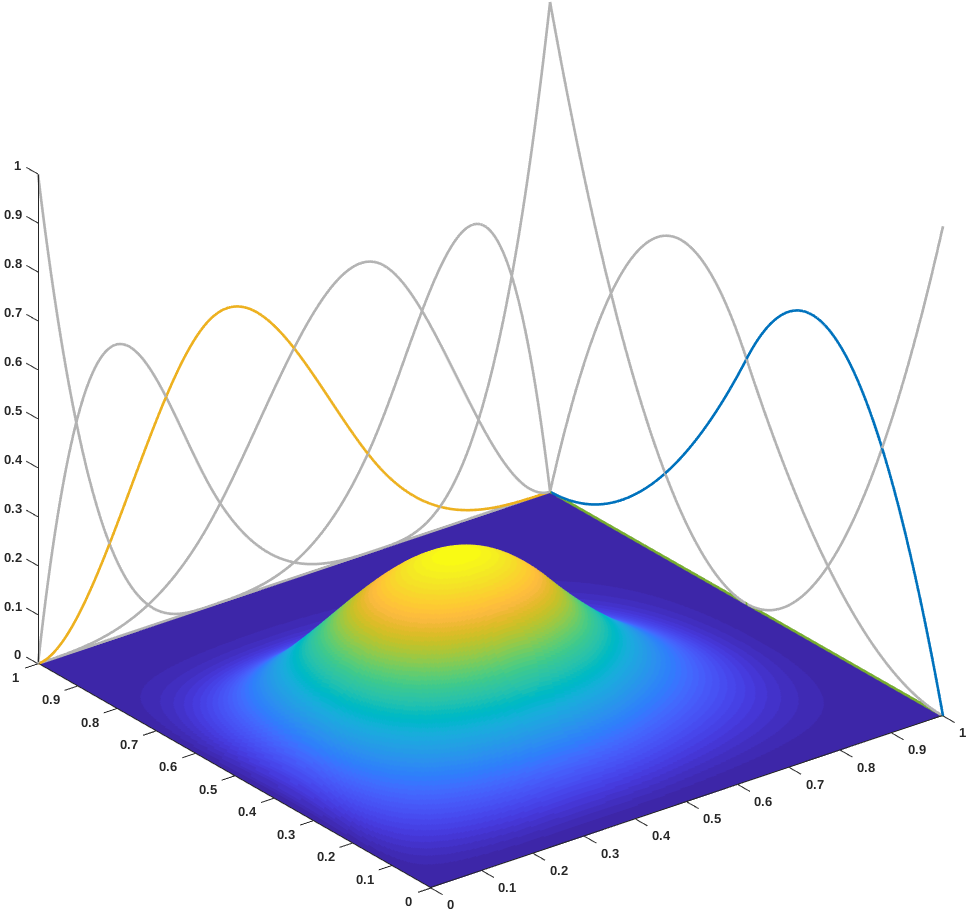}\label{fig:Basis_Bspline_2D}\hfill
   \caption{}
 \end{subfigure}\hspace{5pt}
 \begin{subfigure}[b] {0.47\textwidth}
   \centering
   \includegraphics[width=\textwidth]{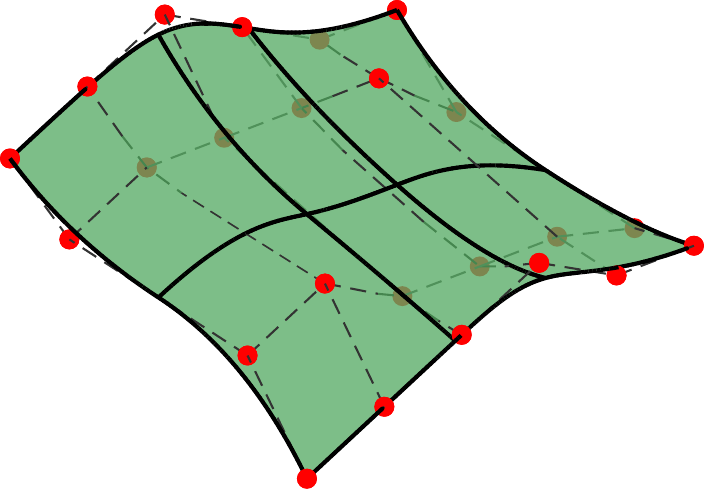}\label{fig:NURBS_2d}\hfill
   \caption{}
 \end{subfigure}\hspace{5pt}
\caption{a) Tensor product basis functions in 2D, cubic B-splines with knot vector $\bm{\xi} = [0,0,0,0,\frac{1}{3},\frac{2}{3},1,1,1,1]$ are shown in the $u$ direction, whereas quadratic B-splines over $\bm{\eta} = [0,0,0,\frac{1}{2},1,1,1]$ are used in the $v$ direction. b) Mapping into the physical space.}
\label{fig:Bspline_2D}
\end{figure}

The basis $\hat{R}_{i,j}^{\hat{p},\hat{q}}$ are rational bivariate B-Splines expressed by:

\begin{equation}
    \hat{R}_{i,j}^{\hat{p},\hat{q}}(\hat{\xi},\hat{\eta}) = \frac{N_{i,\hat{p}}(\hat{\xi})M_{j,\hat{q}}\hat{\eta})\hat{w}_{i,j}}{\sum_{r=1}^{n}\sum_{s=1}^{m}N_{r,\hat{p}}(\hat{\xi})M_{s,\hat{q}}(\hat{\eta})\hat{w}_{r,s}},
\end{equation}
here,  $N_{i,\hat{p}}$, $M_{j,\hat{q}}$ are the B-Splines basis functions of degree $\hat{p}$, $\hat{q}$ respectively. Fig.~\ref{fig:Bspline_2D} shows 2D B-Spline basis function as a result of applying the tensor product of two 1D B-Spline basis functions, and the physical surface.

\section{Volume representations with the IGES exchange format}
\label{S:3}

Let us consider a 3D boundary representation of an object, from which a tetrahedral mesh is generated to model the interior of the domain. If the control points of the tetrahedra facing the boundary of the domain can accurately represent the original boundary representation, we can use the interior elements to solve the required PDEs for analysis while preserving the initial geometry. In order to make the analysis CAD-compatible, the boundary representation is generated by NURBS basis functions, which are mainly used for CAD modeling and allow the creation of a broad variety of curves, i.e. conic shapes. B\'{e}zier tetrahedra are used to mesh the interior of the domain. They also can model conic surfaces, by means of rational basis, and in many cases can give a reasonably good approximation of NURBS objects. An exchange format, however, is required so the same model can be used by different modeling, analysis and mesh generator software packages from a variety of developers.

In the 1970s, efforts to solve this problems derived in the first version of the IGES format. Later, using IGES as a foundation, the STEP file format was published by the International Organization for Standarization (ISO) as a more extensive standard. Together, IGES and STEP are the most used neutral file formats used in the CAD industry. However, other formats developed by CAD software companies are also used as exchange formats, for instance, STL is nowadays broadly used in the 3D printing community.

In this work, we have used the IGES format, although other standards (such as STEP) can be used analogously. IGES, Initial (initially Interim) Graphics Exchange Specification, was the first standard created for the exchange of models between different systems. IGES supports the translation of geometric, topological and non-geometric information. Each basic design component is included into the exchange format as an entity. An entity can contain geometric or non-geometric information. Geometry entities contain information to describe points, curves, surfaces, etc. and include also the relation or connectivity with other geometric entities. Additionally, non-geometric entities, mainly used for 2D drawings, include annotations or attributes such as color.

\begin{figure}[t]
   \centering
   \includegraphics[width=0.8\textwidth]{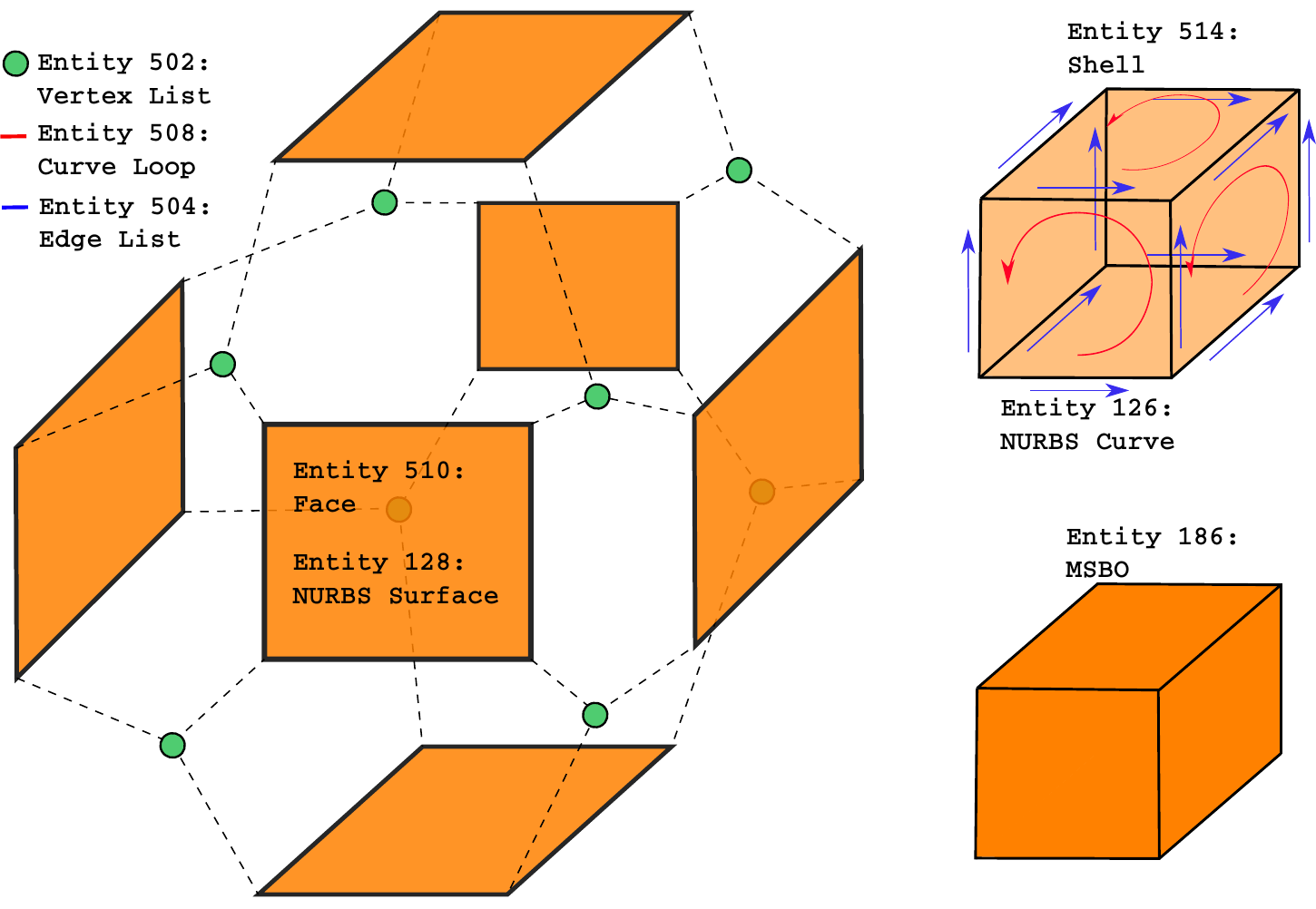}
    \caption{Entity types used to define the MSBO.}
    \label{fig:IGESSolid}
\end{figure}

The construction of NURBS volume domains is supported in IGES through the use of the Manifold Boundary Solid Object (MBSO) entity. This entity defines a volume by specifying the boundary of the 3D object. The boundary is first defined by a set of shells, a connected boundary of a region in $\mathbb{R}^3$. Shells in turn, are decomposed into faces. Each face must point to a surface and a curve loop. The latter is provided in an anticlockwise direction with respect to the outer face of the underlying surface. Finally, the curves and vertices that bound the faces are listed. Fig.~\ref{fig:IGESSolid} shows the construction of a volume domain using the MBSO entity with the IGES exchange format.

\section{Surface reconstruction}
\label{S:4}

In the previous section, NURBS surfaces were used for a boundary representation of a solid object. The next step is to model the interior of the domain with a B\'{e}zier tetrahedral mesh. In this work, we used Gmsh to obtain the finite element mesh, where the tetrahedral elements interpolate the boundaries using Lagrange polynomial based elements. We assume that the boundaries of the Lagrange elements approximate the original NURBS surfaces with control points located exactly on the initial CAD description due to the nodal interpolation property of the Lagrange polynomials. In order to use B\'{e}zier elements instead, their control point coordinates are computed based on the control net of the boundary NURBS surfaces.

A NURBS surface created by tensor product of curves of degree $p$ can be exactly redefined with B\'{e}zier triangles of degree $2p$ \cite{Goldman1987}, which might greatly increase the number of degrees of freedom for the analysis phase. Alternatively, a NURBS surface can be related to B\'{e}zier elements by using B\'{e}zier extraction \cite{Borden2011} and B\'{e}zier projection \cite{Thomas2015}, as shown in \cite{Engvall2017}. However, these are local operators and require each B\'{e}zier element to be located in only one knot span of the NURBS surface. Here, we present a method to link the elements provided by the mesh generator with the original boundary description, regardless of whether the B\'{e}zier element faces are located in one NURBS knot span or not.

First, the control points located at the boundaries of the Lagrange tetrahedral elements are selected. Then, for each tetrahedron, the NURBS parameter space coordinates $\hat{\xi}^e$, $\hat{\eta}^e$ are computed. The mapping of these points into the physical space is thus defined as a linear combination of the shape functions at the computed parameter points and the NURBS control points. For each tetrahedra at the boundary, these physical points should match with the mapping of the corresponding boundary face of the parent B\'{e}zier element into the physical space. For one tetrahedron, this relation is described as:

\begin{equation}\label{eq:physic_point}
	\hat{\bm{x}}\mid_{\Omega^e} = \sum_{i=1}^n N_{i,p}(\hat{\xi}^e,\hat{\eta}^e)\hat{P}_i = \sum_{i=1}^{n_f} B_{i,p}(\hat{\lambda})P_i,
\end{equation}
where $n$ is the number of basis functions of the corresponding NURBS surface and $n_f$ is the number of control points on one face of a B\'{e}zier element with corresponding barycentric coordinates $\bm{\hat{\lambda}}$ in the parent element. In matrix form, Eq.~\eqref{eq:physic_point} can be written as:

\begin{equation}
	\hat{\bm{x}}^e = \bm{N}^e\bm{\hat{P}} = \bm{B}(\hat{\lambda})\bm{P},
\end{equation}
with the matrices of basis functions defined as:
\begin{equation}
	\bm{N}^e = \begin{bmatrix}
	N_{1,p}(\hat{\xi}^e_1,\hat{\eta}^e_1)			&N_{2,p}(\hat{\xi}^e_1,\hat{\eta}^e_1)			&\cdots	&N_{n,p}(\hat{\xi}^e_1,\hat{\eta}^e_1) \\
	N_{1,p}(\hat{\xi}^e_2,\hat{\eta}^e_2)			&N_{2,p}(\hat{\xi}^e_2,\hat{\eta}^e_2)			&\cdots	&N_{n,p}(\hat{\xi}^e_2,\hat{\eta}^e_2) \\
	\vdots										&\vdots										&				&\vdots								\\
	N_{1,p}(\hat{\xi}^e_{n_f},\hat{\eta}^e_{n_f})	&N_{2,p}(\hat{\xi}^e_{n_f},\hat{\eta}^e_{n_f})	&\cdots	&N_{n,p}(\hat{\xi}^e_{n_f},\hat{\eta}^e_{n_f})
	\end{bmatrix},
\end{equation}

\begin{equation}
	\bm{B}(\bm{\hat{\lambda}}) = \begin{bmatrix}
	B_{1,p}(\hat{\lambda}_1)			&B_{2,p}(\hat{\lambda}_1)			&\cdots	&B_{n_f,p}(\hat{\lambda}_1) \\
	B_{1,p}(\hat{\lambda}_2)			&B_{2,p}(\hat{\lambda}_2)			&\cdots	&B_{n_f,p}(\hat{\lambda}_2) \\
	\vdots												&\vdots												&				&\vdots										\\
	B_{1,p}(\hat{\lambda}_{n_f})	&B_{2,p}(\hat{\lambda}_{n_f})	&\cdots	&B_{n_f,p}(\hat{\lambda}_{n_f})
	\end{bmatrix}.
\end{equation}

Thus, the location of the B\'{e}zier control points of the face at the boundary are obtained by inverting the matrix containing the basis functions values at the corresponding parameter points:
\begin{equation}\label{eq:Bezier_reconstruction}
	\bm{P} = \bm{B}^{-1}(\bm{\hat{\lambda}})\bm{N}^e\bm{\hat{P}}.
\end{equation}

With this approach, it is also possible to recover the geometry in cases where the weights are different than one. Defining the coordinates in the projected space $\mathbb{R}^4$ by $ \bm{\tilde{P}} = [xw,yw,zw,w]$, we obtain:
\begin{equation}\label{eq:Bezier_projected_points}
	\bm{\tilde{P}} = \bm{B}^{-1}(\bm{\hat{\lambda}})\bm{N}^e\bm{\hat{\tilde{P}}}.
\end{equation}

\begin{figure}[t]
    \centering
    \includegraphics[width=0.8\textwidth]{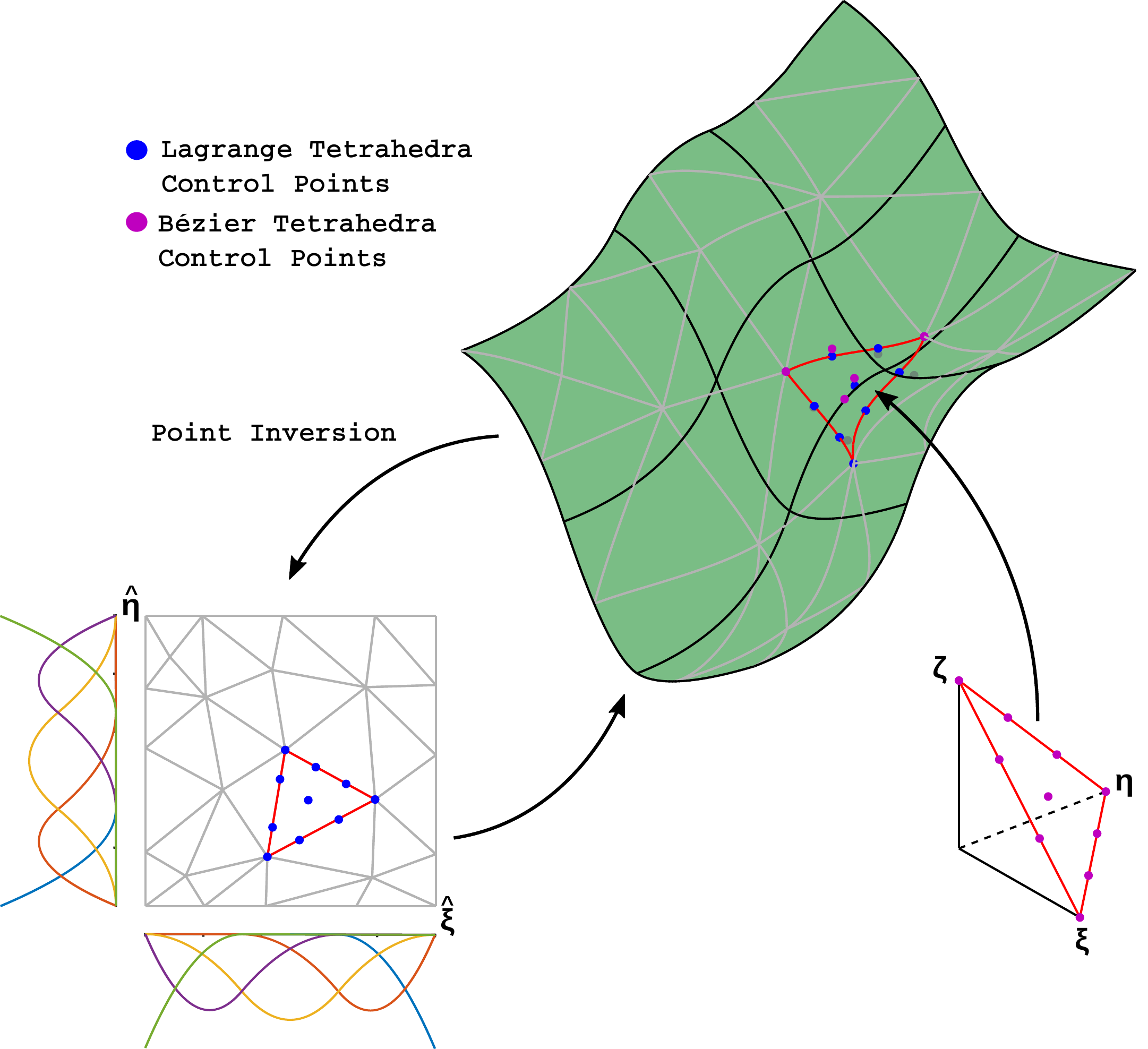}
    \caption{Reconstruction of a NURBS surface by computing the boundary control points of B\'{e}zier tetrahedra.}
    \label{fig:Maping_Srf_Tet}
\end{figure}

The entire procedure to reconstruct the NURBS surface from the given Lagrange triangular mesh is summarized graphically in Fig.~\ref{fig:Maping_Srf_Tet}.

\subsection{Weight smoothing}
The implementation of Eq.~\eqref{eq:Bezier_projected_points} results in control points with variable weights over the surface of the domain. However, it has been proven that the variation of the location of the control points should be minimized in order to achieve optimal convergence \cite{Bazilevs2006,Michoski2016}. In the context of rational polynomials, it means that also the variation of the weights has to be minimized. This is achieved by solving the Laplace equation over the domain for the weights considering the weights at the boundary as essential boundary conditions. The weak formulation of the Laplace problem reads: Find $w\in\mathcal{W} =\{w | w \in (H^1(\Omega))^3, w=\bar{w} \; \text{on } \Gamma\}$ such that for all $\mu \in \mathcal{W}_0 = \{\mu | \mu \in (H^1(\Omega))^3, \mu=0 \; \text{on } \Gamma\}$:

\begin{equation}
    a_w(w,\mu) = 0,
\end{equation}
with the bilinear $a(\cdot)$ operator:
\begin{equation}\label{eq:Smoothing}
        a_w(w,\mu) = \int_\Omega \nabla w\cdot \nabla \mu \,d\Omega.
\end{equation}

Discretizing the continuous functions $w$ and $\mu$ with the same shape functions for each tetrahedron, the discrete weak form reads:
\begin{equation}
    a_w(w^h,\mu^h) = 0,
\end{equation}
where $w^h$ and $\mu^h$ are the discrete trial and test funstions.
Introducing the discretization into Eq.~\eqref{eq:Smoothing}, we get the linear system of equations:
\begin{equation}
    \bm{K}_w \bm{W} = \bm{F}_w,
\end{equation}
where $\bm{F}_w$ is a fictitious force vector, $\bm{W}$ is the vector of weights. The element stiffness matrix $\bm{K}^e_w$ for the Laplace problem is defined as:

\begin{equation}
    \bm{K}^e_w = \int_{\Omega^e}\bm{B}^T_w\bm{I}_3\bm{B}_w \left|\bm{J}\right|d\Omega^e,
\end{equation}
with $\bm{I}_3$ being the 3$\times$3 identity matrix and $\bm{J}$ the Jacobian describing the mapping from the parent element to the physical space:
\begin{equation}
    \bm{J} = \begin{bmatrix}
    \frac{\partial x}{\partial \xi} &\frac{\partial y}{\partial \xi} &\frac{\partial z}{\partial \xi} \vspace{5pt}\\
    \frac{\partial x}{\partial \eta} &\frac{\partial y}{\partial \eta} &\frac{\partial z}{\partial \eta} \vspace{5pt}\\
    \frac{\partial x}{\partial \zeta} &\frac{\partial y}{\partial \zeta} &\frac{\partial z}{\partial \zeta} \vspace{5pt}
    \end{bmatrix}.
\end{equation}

The matrix $\bm{B}_w$ is given by:
\begin{equation}
    \bm{B}_w = \begin{bmatrix}
    \frac{\partial R^p_{p,0,0,0}}{\partial x} &\frac{\partial R^p_{0,p,0,0}}{\partial x} &\cdots &\frac{\partial R^p_{i,j,k,l}}{\partial x} \\
    \frac{\partial R^p_{p,0,0,0}}{\partial y} &\frac{\partial R^p_{0,p,0,0}}{\partial y} &\cdots &\frac{\partial R^p_{i,j,k,l}}{\partial y} \\
    \frac{\partial R^p_{p,0,0,0}}{\partial z} &\frac{\partial R^p_{0,p,0,0}}{\partial z} &\cdots &\frac{\partial R^p_{i,j,k,l}}{\partial z}
    \end{bmatrix}, \hspace{30pt} i+j+k+l=p.
\end{equation}

\section{Shape optimization}
\label{S:5}

Structural shape optimization is a process that seeks for improving the design process by coupling geometrical modeling and analysis of a structure. Together with the use of a proper optimization algorithm, the location of the control points of the CAD model can be found, such that the structural performance is improved. In order to evaluate such characteristics, a typical approach is to minimize the compliance of the system under constraints such as displacements, volume, etc. The statement of the structural shape optimization reads:

\begin{equation}\label{eq:optimization_original}
    \begin{array}{r@{}l}
        &\mbox{minimize} \hspace{11pt} f_0(\bm{s}),\\
        &\mbox{such that} \hspace{10pt} f_i(\bm{s}) \leq 0, \hspace{50pt} i=1,2,...,m,\\
        &\hspace{58pt} s_j^{min} \leq s_j \leq s_j^{max}, \hspace{9pt} j=1,2,...,n,
    \end{array}
\end{equation}
where $f_0(\bm{s})$, $f_i(\bm{s})$ are the objective and constraints functions respectively. In this work, we consider the NURBS control points coordinates of the boundary description as design variables $\bm{s}$. A standard Galerkin discretization is used in order to solve the linear elasticity problem and evaluate the objective and constraints functions. The weak form of the linear elasticity equation is: Find $\bm{u}\in\mathcal{U} =\{\bm{u} | \bm{u} \in (H^1(\Omega))^3, \bm{u}=\bm{\bar{u}} \; \text{on } \Gamma_u\}$ such that for all $\bm{\psi} \in \mathbf{\Psi} = \{\bm{\psi} | \bm{\psi} \in (H^1(\Omega))^3, \bm{\psi}=\bm{0} \; \text{on } \Gamma_u\}$:
   
\begin{equation}\label{eq:Linear_Elasticity}
     a_u(\bm{u},\bm{\psi})=l_u(\bm{\psi}),
\end{equation}
with:
\begin{equation}\label{eq:Linear_Elasticity2}
     a_u(\bm{u},\bm{\psi}) = \int_{\Omega} \bm{\epsilon}(\bm{u}):\mathbb{C}:\bm{\epsilon}(\bm{\psi})d\Omega,
\end{equation}
and
\begin{equation}\label{eq:Linear_Elasticity3}
    l_u(\bm{\psi})=\int_{\Omega} \bm{b}\cdot\bm{\psi}d\Omega + \int_{\Gamma^N} \bm{t}\cdot\bm{\psi}d\Gamma^N,
\end{equation}
where $\bm{\epsilon}$ denotes the strain, $\mathbb{C}$ is the fourth-order elasticity tensor, $\bm{b}$ is the body force over the domain and $\bm{t}$ is the traction over the boundary $\Gamma^N$. Using the isoparametric concept the displacement vector $\bm{u}$, and the test function $\bm{\psi}$ are discretized using the same basis functions on an element $\Omega^e$. Then, the weak formulation can be written in discrete form as:
\begin{equation}\label{eq:Linear_Elasticity4}
     a_u(\bm{u}^h,\bm{\psi}^h)=l_u(\bm{\psi}^h),
\end{equation}
where $\bm{u}^h$, $\bm{\psi}^h$ are the semidiscrete displacement and test functions respectively. Using these relations and \eqref{eq:Linear_Elasticity4} for each tetrahedron, we can obtain the discrete form of the balance equation:
\begin{equation}\label{eq:elasticity_equilibrum}
    \bm{K}_u\bm{U} = \bm{t},
\end{equation}
where $\bm{U}$ is the vector containing the discrete values $u_{ijkl}$, and $\bm{K}_u$ is the stiffness matrix for the elasticity problem, defined for each tetrahedron as:
\begin{align}
    \bm{K}^e_u&=\int_{\hat{\Omega}_e} \bm{B}_u^{T}\bm{C}\bm{B}_u\left|\bm{J}\right|d\hat{\Omega}_e,
\end{align}
here $\bm{C}$ is the constitutive matrix, and the matrix $\bm{B}_u$ is described by:
\begin{equation}\label{eq:Bu}
    \bm{B}_u = \begin{bmatrix}
    \frac{\partial R^p_{p,0,0,0}}{\partial x} &0 &0 &\cdots &\frac{\partial R^p_{i,j,k,l}}{\partial x} &0 &0\\
    0 &\frac{\partial R^p_{p,0,0,0}}{\partial y} &0 &\cdots &0 &\frac{\partial R^p_{i,j,k,l}}{\partial y} &0\\
    0 &0 &\frac{\partial R^p_{p,0,0,0}}{\partial z}  &\cdots &0 &0 &\frac{\partial R^p_{i,j,k,l}}{\partial z}\\
    \frac{\partial R^p_{p,0,0,0}}{\partial y} &\frac{\partial R^p_{p,0,0,0}}{\partial x} &0 &\cdots &\frac{\partial R^p_{i,j,k,l}}{\partial y} &\frac{\partial R^p_{i,j,k,l}}{\partial x} &0\\
    0 &\frac{\partial R^p_{p,0,0,0}}{\partial z} &\frac{\partial R^p_{p,0,0,0}}{\partial y} &\cdots &0 &\frac{\partial R^p_{i,j,k,l}}{\partial z} &\frac{\partial R^p_{i,j,k,l}}{\partial y}\\
    \frac{\partial R^p_{p,0,0,0}}{\partial z} &0 &\frac{\partial R^p_{p,0,0,0}}{\partial x} &\cdots &\frac{\partial R^p_{i,j,k,l}}{\partial z} &0 &\frac{\partial R^p_{i,j,k,l}}{\partial x}
    \end{bmatrix}, \hspace{10pt} i+j+k+l=p.
\end{equation}

\subsection{Sensitivity analysis}
\label{S:5.1}
We use a gradient based optimization algorithm (MMA), therefore the derivatives of the objective and constraints functions with respect to the design variables are required. In this work, we consider displacement and compliance minimization problems. For both cases, the gradient $d\bm{u}/ds$ needs to be computed. In the following, we consider $s$ to be one entry of the vector of design variables $\bm{s}$. The derivative of the displacement vector with respect to the design variables can be obtained from the equilibrium equation~\eqref{eq:elasticity_equilibrum}:
\begin{align}
\bm{U}_{,s}&=\bm{K}^{-1}\left(\bm{f}_{,s}-\bm{K}_{,s}\right),
\end{align}
where $\bm{U}_{,s}$, $\bm{f}_{,s}$ are the gradients of the displacement and force vectors with respect to the design variables respectively. The stiffness matrix gradient $\bm{K},s$ is defined for each element as:
\begin{align}
\bm{K}_{e,s}&=\int_{\hat{\Omega}_e}\left(\bm{B}^T_{u,s}\bm{C}\bm{B}_u\left|\bm{J}\right| + \bm{B}_u^T\bm{C}\bm{B}_{u,s}\left|\bm{J}\right| + \bm{B}_u^T\bm{C}\bm{B}_u\left|\bm{J}\right|_{,s}\right)d\hat{\Omega}_e,
\end{align}
with the gradient $\left|\bm{J}\right|_{,s}$ obtained by:
\begin{equation}
\left|\bm{J}\right|_{,s} = \left|\bm{J}\right|\mbox{tr}\left(\bm{J}^{-1}\bm{J}_{,s}\right).
\end{equation}

The sensitivity of the Jacobian is computed with:
\begin{equation}
    \bm{J}_{,s} = \begin{bmatrix}
    \frac{\partial x_{,s}}{\partial \xi} &\frac{\partial y_{,s}}{\partial \xi} &\frac{\partial z_{,s}}{\partial \xi} \vspace{5pt}\\
    \frac{\partial x_{,s}}{\partial \eta} &\frac{\partial y_{,s}}{\partial \eta} &\frac{\partial z_{,s}}{\partial \eta} \vspace{5pt}\\
    \frac{\partial x_{,s}}{\partial \zeta} &\frac{\partial y_{,s}}{\partial \zeta} &\frac{\partial z_{,s}}{\partial \zeta} \vspace{5pt}
    \end{bmatrix}.
\end{equation}

The gradient $\bm{B}_{u,s}$ can be computed as follows, and rearranging the entries as in \eqref{eq:Bu}:
\begin{equation}
\begin{bmatrix}\frac{\partial R_{ijkl}}{\partial x}\\[6pt] \frac{\partial R_{ijkl}}{\partial y}\\[6pt] \frac{\partial R_{ijkl}}{\partial z} \end{bmatrix} = \bm{J}^{-1}_{,s} \begin{bmatrix}\frac{\partial R_{ijkl}}{\partial \xi}\\[6pt] \frac{\partial R_{ijkl}}{\partial \eta} \\[6pt] \frac{\partial R_{ijkl}}{\partial \zeta}\end{bmatrix},
\end{equation}
with
\begin{equation}
\bm{J}^{-1}_{,s}= -\bm{J}^{-1}\bm{J}_{,s}\bm{J}^{-1},
\end{equation}

Recalling \eqref{eq:Bezier_reconstruction}  and \eqref{eq:Bezier_projected_points}, the sensitivity $\bm{J}_{,s}$ can be written in matrix form as:

\begin{equation}
    \bm{J}_{,s} = \bm{B}_{,\bm{\xi}}\hat{\bm{B}}^{-1}(\hat{\lambda})\hat{\bm{N}}^e\bm{\hat{P}_{,s}},
\end{equation}
with
\begin{equation}
    \bm{B}_{,\bm{\xi}} = \begin{bmatrix}
    \frac{\partial R^p_{p,0,0,0}}{\partial \xi} &\frac{\partial R^p_{0,p,0,0}}{\partial \xi} &\cdots &\frac{\partial R^p_{i,j,k,l}}{\partial \xi} \vspace{5pt}\\
    \frac{\partial R^p_{p,0,0,0}}{\partial \eta} &\frac{\partial R^p_{0,p,0,0}}{\partial \eta} &\cdots &\frac{\partial R^p_{i,j,k,l}}{\partial \eta} \vspace{5pt}\\
    \frac{\partial R^p_{p,0,0,0}}{\partial \zeta} &\frac{\partial R^p_{0,p,0,0}}{\partial \zeta} &\cdots &\frac{\partial R^p_{i,j,k,l}}{\partial \zeta}
    \end{bmatrix}, \hspace{30pt} i+j+k+l=p,
\end{equation}
and
\begin{equation}
    \bm{\hat{P}}_{,s} = \begin{bmatrix}
    \hat{P}^{x}_{1,s} &\hat{P}^{y}_{1,s} &\hat{P}^{z}_{1,s} \vspace{5pt}\\
    \hat{P}^{x}_{2,s} &\hat{P}^{y}_{2,s} &\hat{P}^{z}_{2,s} \vspace{5pt}\\
    \vdots &\vdots &\vdots \vspace{5pt}\\
    \hat{P}^{x}_{n,s} &\hat{P}^{y}_{n,s} &\hat{P}^{z}_{n,s} \vspace{5pt}\\
    \end{bmatrix},
\end{equation}

As mentioned before, we consider the design variables to be one of the coordinates of the NURBS control points $s=\hat{P}^{i}_{j}\cdot e_i$. Hence, we can compute the sensitivities $\hat{P}^{\alpha}_{\beta,s}$ by:
\begin{equation}
    \frac{\partial\hat{P}^{\alpha}_{\beta}}{\partial\hat{P}^{i}_{j}} = \begin{cases} 1 \hspace{5pt} \mbox{if} \hspace{5pt} \alpha = i \hspace{5pt} \mbox{and} \hspace{5pt} \beta = j, \\
    0 \hspace{5pt} \mbox{otherwise},
    \end{cases}
\end{equation}
with $\alpha =$ 1, 2, 3 corresponding to the $x$, $y$, $z$ directions.  

\begin{figure}[t]
   \centering
   \includegraphics[width=0.6\textwidth]{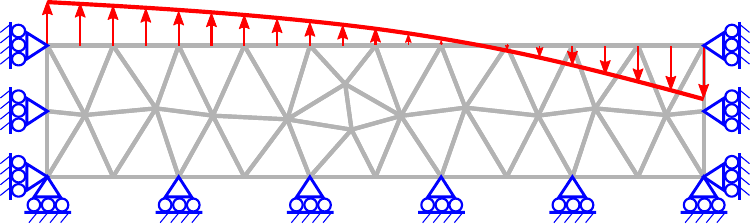}
\caption{Boundary conditions for the pseudo-elastic problem applied for the mesh moving algorithm on a 2D cantilever beam.}
\label{fig:PseudoBC}
\end{figure}

\subsection{Method of moving asymptotes}
The method of moving asymptotes (MMA) \cite{Svanberg1987} is a non-linear optimization algorithm within the Conservative Convex Separable Approximations (CCSA) methods. These methods are characterized by solving a sequence of convex subproblems at each iteration, which are approximations of the original problem. In general, the MMA solves the problem statement from \eqref{eq:optimization_original} by employing the following formulation:

\begin{equation}\label{eq:optimization_2}
    \begin{array}{r@{}l}
        &\mbox{minimize} \hspace{11pt} g_0^{(k)}(\bm{s}) + a_0z + \sum_{i=1}^m(c_iy_i + \frac{1}{2}d_iy_i^2),\\
        &\mbox{such that} \hspace{10pt} g_i^{(k)}(\bm{s}) -a_iz -y_i \leq 0, \hspace{80pt} i=1,2,...,m,\\
        &\hspace{58pt} \hat{\alpha}_j^{(k)} \leq s_j \leq \hat{\beta}_j^{(k)}, \hspace{107pt} j=1,2,...,n,\\
        &\hspace{58pt} z \geq 0 \hspace{7pt} \mbox{and} \hspace{7pt} y_i \geq 0 \hspace{102pt} i=1,2,...,m,
    \end{array}
\end{equation}
where $y_i$, $z$ are artificial variables included to ensure the feasibility of the problem. To approximate the original problem from \eqref{eq:optimization_original}, the values for the constants are set $a_0=1$, $a_i=0$, $c_i=1000$, $d_i=1$ for all $i$. The approximation functions $g_i$ used in the MMA are:

\begin{equation}\label{eq:MMA_Approx}
    g_i^{(k)} = \sum_{j=1}^{n}\left( \frac{p_{ij}^{(k)}}{u_j^{(k)}-s_j^{(k)}} + \frac{q_{ij}^{(k)}}{s_j^{(k)}-l_j^{(k)}}\right), \hspace{25pt} i=0,1,...,m,
\end{equation}
with:
\begin{equation}
    p_{ij}^{(k)} = (u_j^{(k)}-s_j^{(k)})^2\Bigg(0.001 \hspace{2pt}\mbox{max} \bigg\{\frac{\partial f_i}{\partial s_j}(\bm{s}^{(k)},0)\bigg\} + \frac{10^{-5}}{s_j^{max}-s_j^{min}}  \Bigg),
\end{equation}

\begin{equation}
    q_{ij}^{(k)} = (s_j^{(k)}-l_j^{(k)})^2\Bigg(0.001 \hspace{2pt}\mbox{min} \bigg\{-\frac{\partial f_i}{\partial s_j}(\bm{s}^{(k)},0)\bigg\} + \frac{10^{-5}}{s_j^{max}-s_j^{min}}  \Bigg),
\end{equation}
where the superscript $(k)$ denotes the iteration number. The design variable boundaries $\alpha_j^{(k)}$, $\beta_j^{(k)}$ used for the approximation problem are computed by:
\begin{equation}
\begin{array}{r@{}l}
\alpha_j^{(k)} &= \mbox{max}\{s_j^{\mbox{min}},l_j^{(k)} + \theta (s_j^{(k)}-l_j^{(k)})\}, \\
\beta_j^{(k)} &= \mbox{min}\{s_j^{\mbox{max}},u_j^{(k)} + \theta (u_j^{(k)}-s_j^{(k)})\},
\end{array}
\end{equation}
where $l_j^{(k)}$, $u_j^{(k)}$ are the lower and upper asymptotes of the convex approximations. The distance between the design variables and their corresponding $\alpha_j^{(k)}$, $\beta_j^{(k)}$ bounds is restrained with the factor $\theta$. The lower and upper asymptotes are computed for the first two iterations as:

\begin{equation}
\begin{array}{r@{}l}
l_j^{(k)} &= s_j^{(k)} - \gamma^{(0)}(s_j^{\mbox{max}} - s_j^{\mbox{min}}), \\
u_j^{(k)} &= s_j^{(k)} + \gamma^{(0)}(s_j^{\mbox{max}} - s_j^{\mbox{min}}),
\end{array}
\end{equation}
and for $k\geq 3$ by:
\begin{equation}
\begin{array}{r@{}l}
l_j^{(k)} &= s_j^{(k)} - \gamma^{(k)}(s_j^{(k-1)} - l_j^{(k-1)}), \\
u_j^{(k)} &= s_j^{(k)} + \gamma^{(k)}(u_j^{(k-1)} - s_j^{(k-1)}),
\end{array}
\end{equation}
with the variable $\gamma^{(k)}$ as:
\begin{equation}
\gamma^{(k)} = \begin{cases}
\gamma_\ell \hspace{6pt} \mbox{if} \hspace{3pt} (s_j^{(k)} - s_j^{(k-1)} )(s_j^{(k-1)} - s_j^{(k-2)}) < 0, \\
\gamma_u \hspace{8pt} \mbox{if} \hspace{3pt} (s_j^{(k)} - s_j^{(k-1)} )(s_j^{(k-1)} - s_j^{(k-2)}) > 0, \\
1 \hspace{12pt} \mbox{if} \hspace{3pt} (s_j^{(k)} - s_j^{(k-1)} )(s_j^{(k-1)} - s_j^{(k-2)}) = 0,
\end{cases}
\end{equation}
here, $\gamma_\ell$, $\gamma_u$ are constants that control the movement of the asymptotes depending on the direction the design variables are taking.

\subsection{Mesh update}
\label{S:5.2}

During the optimization process, the control points chosen as design variables can move according to the objective, constraints and algorithm. Therefore, an appropriate method to update the mesh is needed. A simple approach is to regenerate the mesh at each iteration after changing the boundary of the geometry. However, depending on the size of the problem, this might lead to an increase of the computational time. Moreover, the mesh topology changes can also cause a non-monotonic optimization convergence.

\begin{algorithm}[h]
	\caption{Shape optimization with B\'{e}zier tetratedra and a NURBS boundary description}	
	\begin{tabularx}{\textwidth}{l>{$}c<{$}X} 
	&\textbf{Input:} &\mbox{Set of NURBS surfaces describing the boundary domain} \\ & &\mbox{Initial values of the design variable vector($\bm{s}_0$)} \\ & &\mbox{Minimum and maximum limits for the design variable vector} \\ & &\mbox{Maximum number of iterations} ($it_{max}$)\\ & &\mbox{Termination tolerance $\epsilon$}\\ & &\mbox{Element distortion tolerance $\delta$}\\
	&\textbf{Output:} & Set of NURBS surfaces describing the optimal boundary shape\\& &Optimal design variable vector ($\bm{s}_{opt}$)\\& & Objective function value ($f_0$)\\& &Constraints values ($f_i(\bm{s})$)\\
	\end{tabularx}
	\textbf{Method:}
	\begin{algorithmic}[1]
	    \State{Initial Boundary representation of a CAD solid}
	    \State{Represent the volume with a CAD Exchange format (IGES)}
	    \State{Generate mesh (Gmsh)}
	    \State{Read mesh file: extract nodes coordinates and topology connectivity}
	    \State{Compute B\'{e}zier control points at the boundary based on the NURBS surfaces}
	    \State{Smooth weights}
	   \While{$\left|\frac{f_0^{it} - f_0^{{it}-1}}{f_0^0}\right| > \epsilon$ and $it\leq it_{max}$}

		 \State{Solve PDE: perform analysis and sensitivity analysis}
		 \State{Optimization algorithm (MMA): compute new location of design variables, update the current boundary design}
		 \State{Solve pseudo-elastic problem: compute new location of the mesh nodes}
		 \State{Update geometry to a CAD Exchange format (IGES)}
		 \EndWhile
		\algstore*{Insert}
	\end{algorithmic}
	\label{alg:Optimization}
\end{algorithm}

Body-fitted approaches of moving mesh problems consider the mesh as a deformable body. The problem consists on finding the location of the node locations of the interior of the domain given the displacement of the boundaries. A common method used in fluid-structure interaction and optimization problems is to solve a linear elasticity problem (also called pseudo-elastic problem) with pure Dirichlet boundary conditions. 
Consider the boundary of the domain $\Gamma = \Gamma_d \cup \Gamma_s$. Here, $\Gamma_d$ denotes the design boundary where the design control points are located and a fictitious displacement $\bar{\bm{d}}$ is imposed, while on $\Gamma_s$, the control points are allowed to slip along the boundary. Fig.~\ref{fig:PseudoBC} shows the boundary conditions imposed over a 2D mesh in order to illustrate the mesh moving problem. The pseudo-elastic problem with no body force nor traction is as follows: Find $\bm{d}\in\mathcal{D} =\{\bm{d} | \bm{d} \in (H^1(\Omega))^3, \bm{d}=\bm{\bar{d}} \; \text{on } \Gamma_d\}$ such that for all $\bm{\chi} \in \mathbf{X} = \{\bm{\chi} | \bm{\chi} \in (H^1(\Omega))^3, \bm{\chi}=\bm{0} \; \text{on } \Gamma_d\}$:

\begin{equation}\label{eq:PseudoElasticity}
     a_d(\bm{d},\bm{\chi})=0,
\end{equation}
with:
\begin{equation}
     a_d(\bm{d},\bm{\chi}) = \int_{\Omega} \bm{\epsilon}(\bm{d}):\mathbb{C}_{mesh}:\bm{\epsilon}(\bm{\chi})d\Omega,
\end{equation}
where $\bm{d}$ is the displacement of the control points of the mesh, $\chi$ is the test function and with $\mathbb{C}_{mesh}$ as the fourth-order elastic tensor for the mesh, which is defined for each element in order to control their stiffness and retain mesh quality as much as possible. Similarly as in the linear elasticity problem, the trial and test functions are discretized into $\bm{d}^h$, $\bm{\chi}^h$ using the same basis functions for each tetrahedron to obtain the discrete equation:
\begin{equation}
     a_d(\bm{d}^h,\bm{\chi}^h)=0,
\end{equation}

Again, introducing the discretization equations into Eq.~\eqref{eq:PseudoElasticity}, we can obtained the discrete balance equation for the pseudo-elastic problem:
\begin{equation}\label{eq:elasticMesh}
    \bm{K}_m \bm{D} = \bm{F}_m
\end{equation}
 $\bm{K}_m$ and $\bm{F}_m$ are the fictitious stiffness matrix and force vector respectively, and $\bm{D}$ is the vector of displacements $d_{ijkl}$. 
 As in \cite{Johnson1994}, the same elastic matrix used in the solution of the linear elasticity problem is used as elastic matrix for the mesh, $\bm{C}_{mesh}=\bm{C}$. Also, the Jacobian is dropped from the definition of $\bm{K}_m$, so bigger elements have lower stiffness values as:
 \begin{align}
    \bm{K}^e_m&=\int_{\hat{\Omega}_e} \bm{B}_m^{T}\bm{C}\bm{B}_m d\hat{\Omega}_e,
\end{align}
Finally, the control point coordinates of the tetrahedral mesh at the $t^{th}$ iteration is obtained by adding the displacement vector to the control points location at the previous iteration:
 
 \begin{equation}
    \bm{P}_t = \bm{P}_{t-1} + \bm{D}, \hspace{10pt} t=2,3,...,it_{max}.
\end{equation}

The complete procedure of the structural optimization approach is summarized in Algorithm~\ref{alg:Optimization}.

\begin{figure}[t]
 \centering
 \begin{subfigure}[b] {0.37\textwidth}
   \centering
   \includegraphics[width=\textwidth]{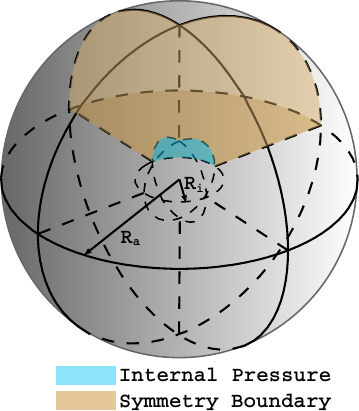}\hfill
   \caption{}
 \end{subfigure}\hspace{15pt}
 \begin{subfigure}[b] {0.45\textwidth}
   \centering
   \includegraphics[width=\textwidth]{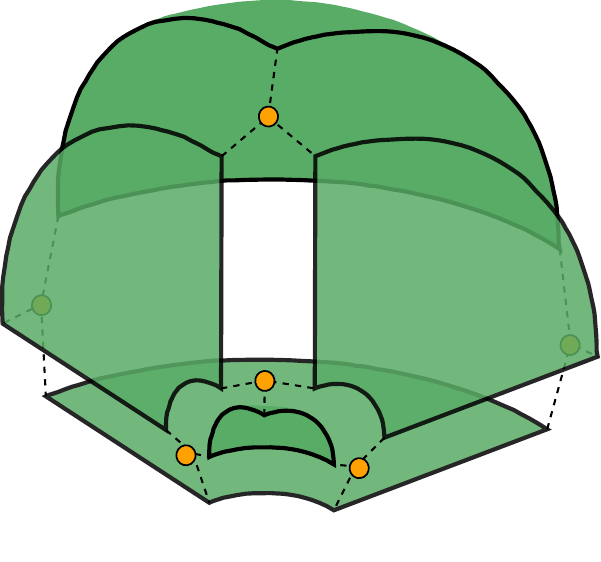}
   \caption{}
 \end{subfigure}
 \caption{a) Dimensions and boundary conditions for the hollow sphere validation example. b) Boundary representation of one eight of the sphere.}
 \label{fig:HollowSphere}
 \end{figure}
 
\section{Numerical examples}
\label{S:6}
In this section, we present some numerical examples that demonstrate the effectiveness of the method. The first example validates the convergence of the B\'{e}zier tetrahedron element developed. Next, numerical examples for the solution of 3D structural shape optimization problems are presented. The results presented were obtained using a dual Intel(R) Xeon(R) CPU E5-2650 v2 @ 2.60GHz processor with 64 Gb RAM. The optimization algorithm was implemented in Matalab 2018b, and the mesh was generated using the Gmsh version 4.2.2. For all the examples, the termination and distortion tolerances are set $\epsilon=10^{-4}$, $\delta=1.8$.

\subsection{Validation test: Pressurized sphere}
 To validate the B\'{e}zier tetrahedron, we consider the problem of a hollow sphere under internal pressure. The sphere has an external radius $R_a = 4$ m and internal radius $R_i = 1$ m, the internal pressure is $P=1$ Pa. The material properties used are the elasticity modulus $E=1000$ Pa, and Possin's ratio $\nu=0.3$. Due to symmetry, only one eighth of the domain is modelled. Fig.~\ref{fig:HollowSphere} shows the dimensions and the imposed boundary conditions. The exact stresses using the spherical coordinate system $[r,\phi,\theta]$ are as follows:
\begin{equation}
    \begin{split}
        \sigma_{rr} & = \frac{PR_{i}^{3}(R_{a}^{3}-r^3)}{r^3(R_{a}^{3}-R_{i}^{3})},\\
        \sigma_{\phi\phi} & = \frac{PR_{i}^{3}(R_{a}^{3}+2r^3)}{2r^3(R_{a}^{3}-R_{i}^{3})},\\
        \sigma_{\theta\theta} & = \sigma_{\phi\phi}.
    \end{split}
\end{equation}

\begin{figure}[t]
   \centering
   \includegraphics[width=0.5\textwidth]{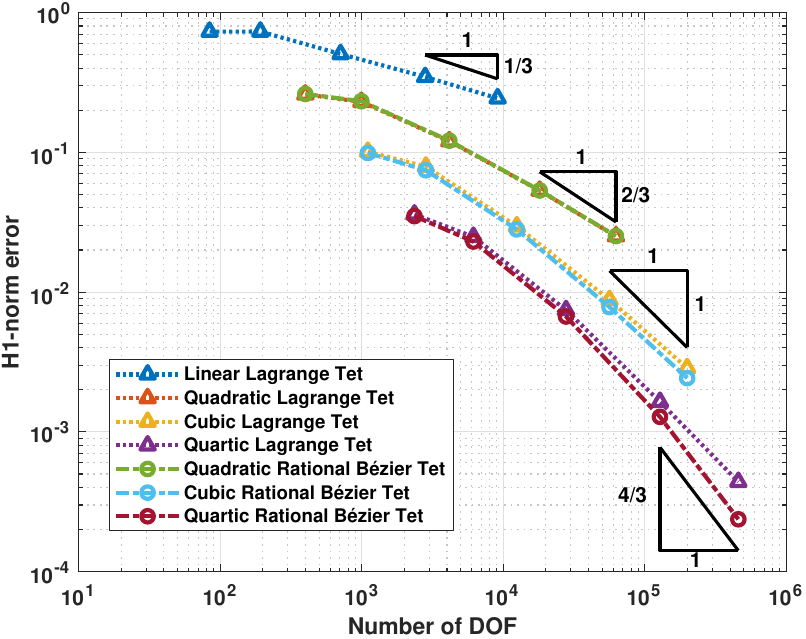}
\caption{Energy norm error convergence for the hollow sphere example.}
\label{fig:H1ErrorSphere}
\end{figure}

Fig.~\ref{fig:H1ErrorSphere} shows the energy norm convergence for Lagrange tetrahedra of degree $p=$1, 2, 3 and 4, as well as for quadratic, cubic and quartic rational B\'{e}zier tetrahedra. The results were obtained by generating a mesh with element sizes of $h_l=$ 4, 2, 1, 0.5, 0.3. With this example, we show that the accuracy of the $C^0$ continuous B\'{e}zier tetrahedra is comparable to the Lagrange elements, which is expected since both bases span the same space of piecewise polynomials of a given degree. However, it can be seen that the $H_1$ norm error is slightly lower for rational B\'{e}zier elements, especially for higher polynomial degrees.
 
\begin{figure}[t]
 \centering
 \begin{subfigure}[t] {0.25\textwidth}
   \centering
   \includegraphics[width=\textwidth]{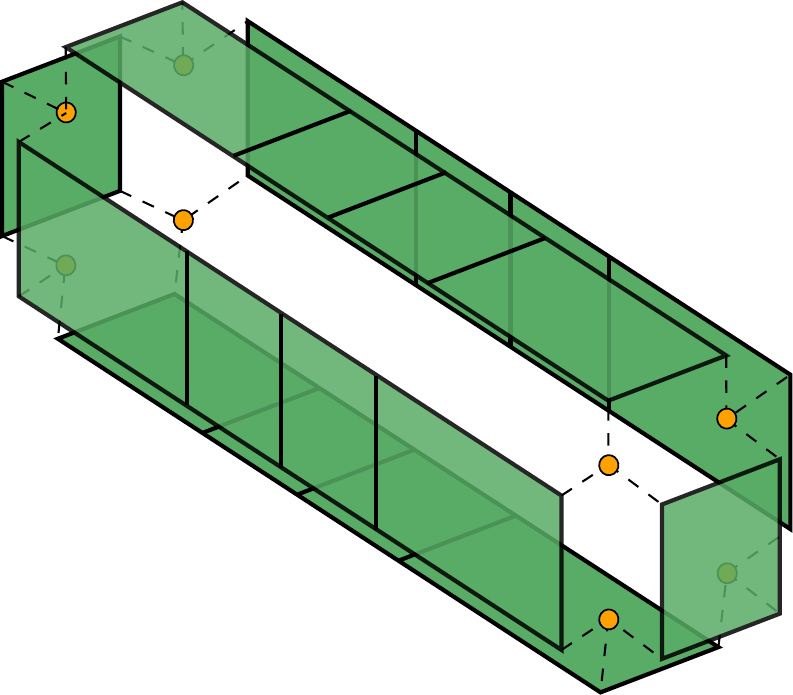}\hfill
   \caption{}
 \end{subfigure}\hspace{5pt}
 \begin{subfigure}[t] {0.25\textwidth}
   \centering
   \includegraphics[width=\textwidth]{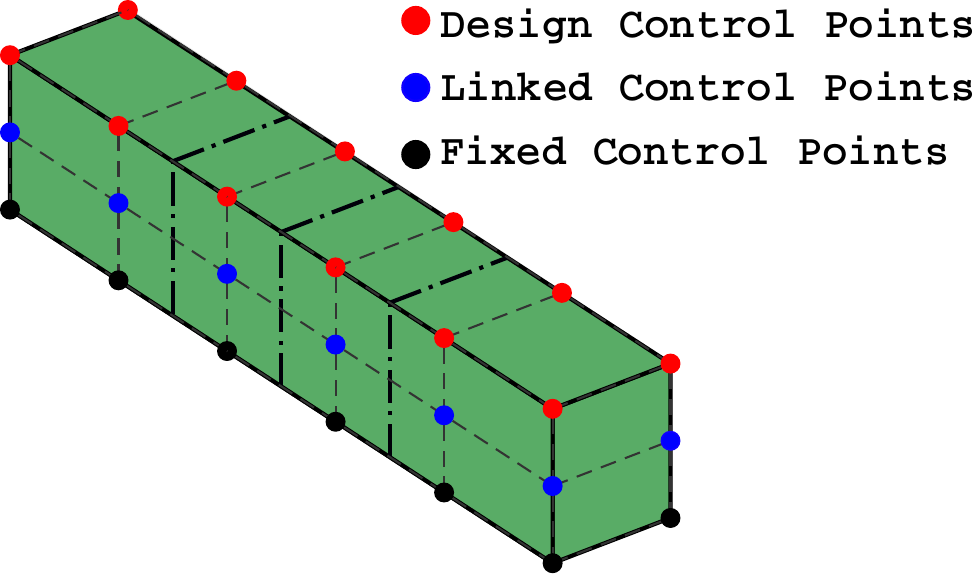}\hfill
   \caption{}
 \end{subfigure}\hspace{5pt}
 \begin{subfigure}[t] {0.33\textwidth}
   \centering
   \includegraphics[width=\textwidth]{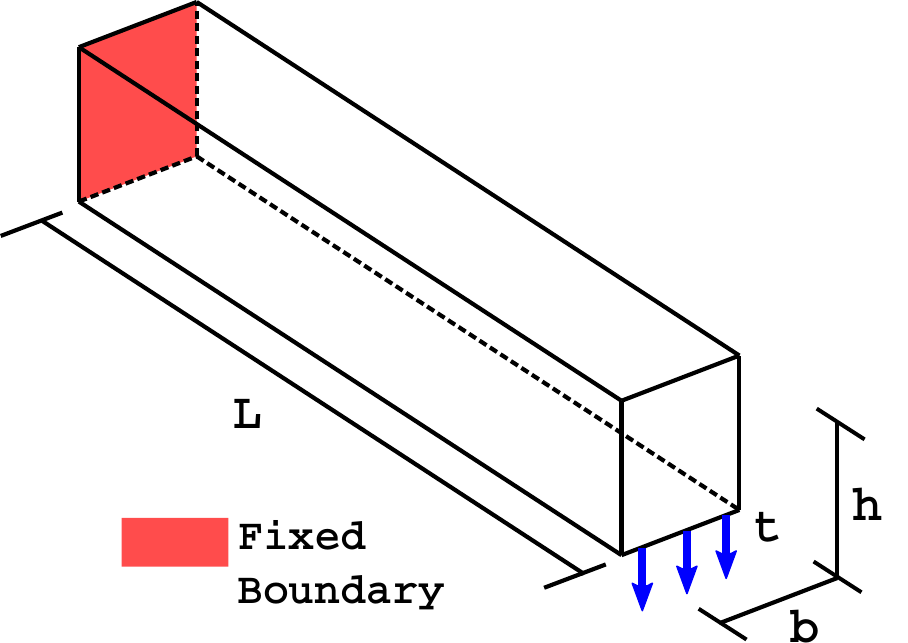}
   \caption{}
 \end{subfigure}
 \caption{a) Initial set of surfaces used to describe the boundary of the cantilever beam. b) Initial location of the design and linked control points. c) Dimensions and boundary conditions for the problem.}
 \label{fig:CantileverInitial}
 \end{figure}
 
 \begin{figure}[H]
 \centering
 \begin{subfigure}[t] {0.3\textwidth}
   \centering
   \includegraphics[width=\textwidth]{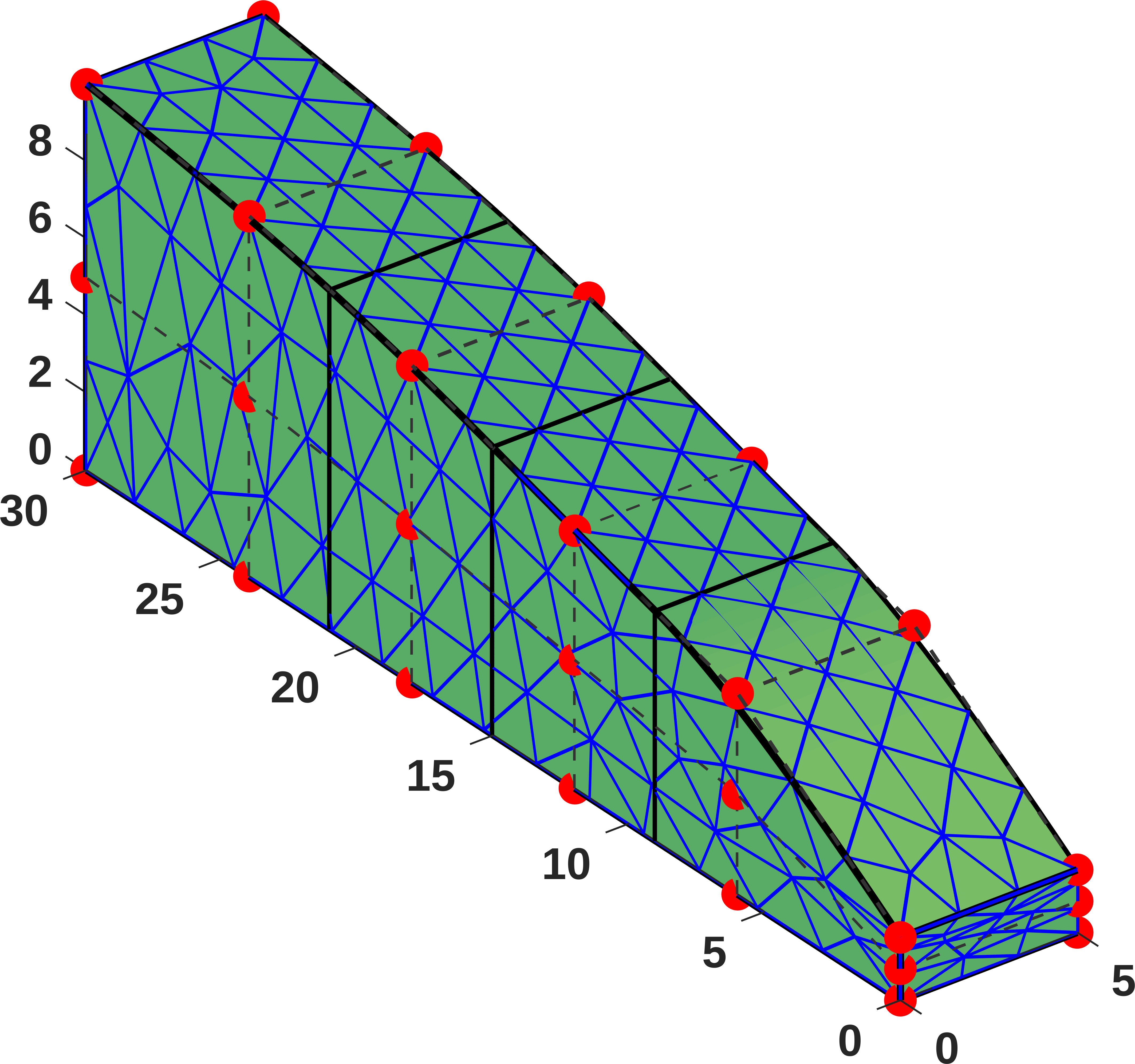}\hfill
   \caption{}
 \end{subfigure}\hspace{15pt}
 \begin{subfigure}[t] {0.4\textwidth}
   \centering
   \includegraphics[width=\textwidth]{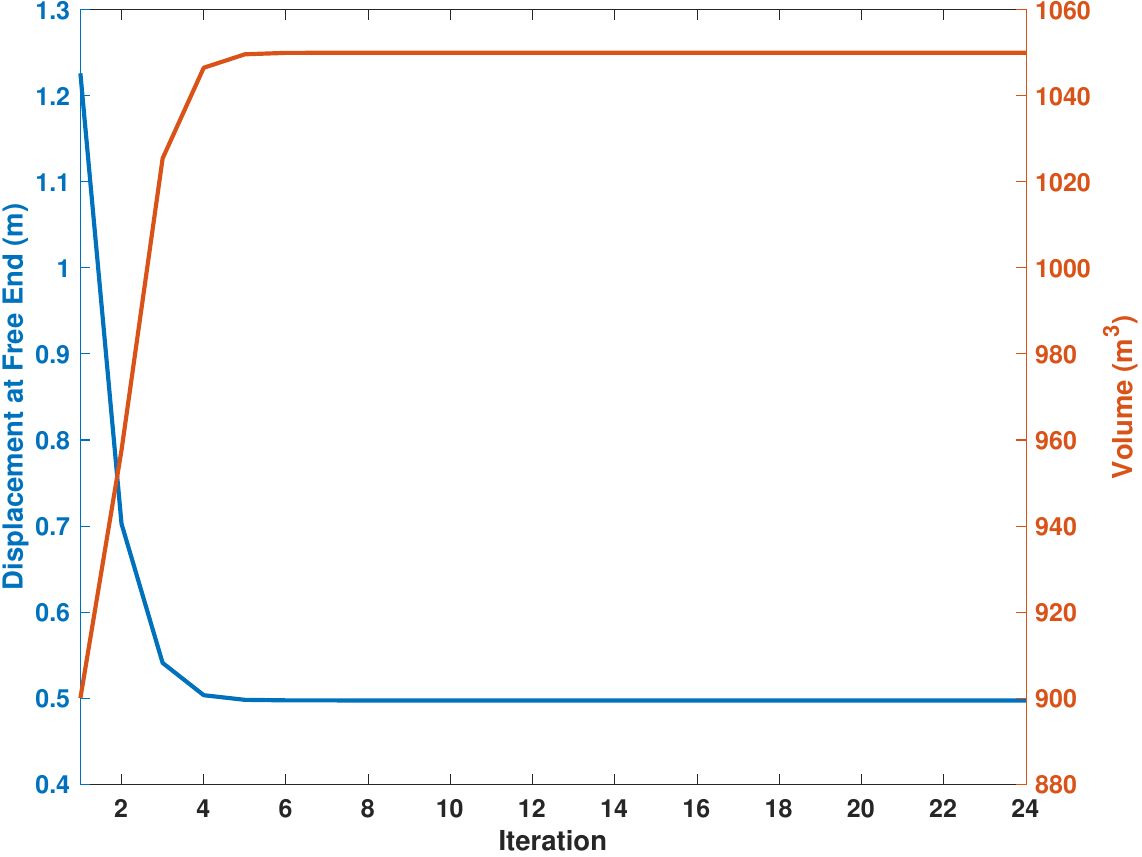}
   \caption{}
 \end{subfigure}
 
 \begin{subfigure}[t] {0.3\textwidth}
   \centering
   \includegraphics[width=\textwidth]{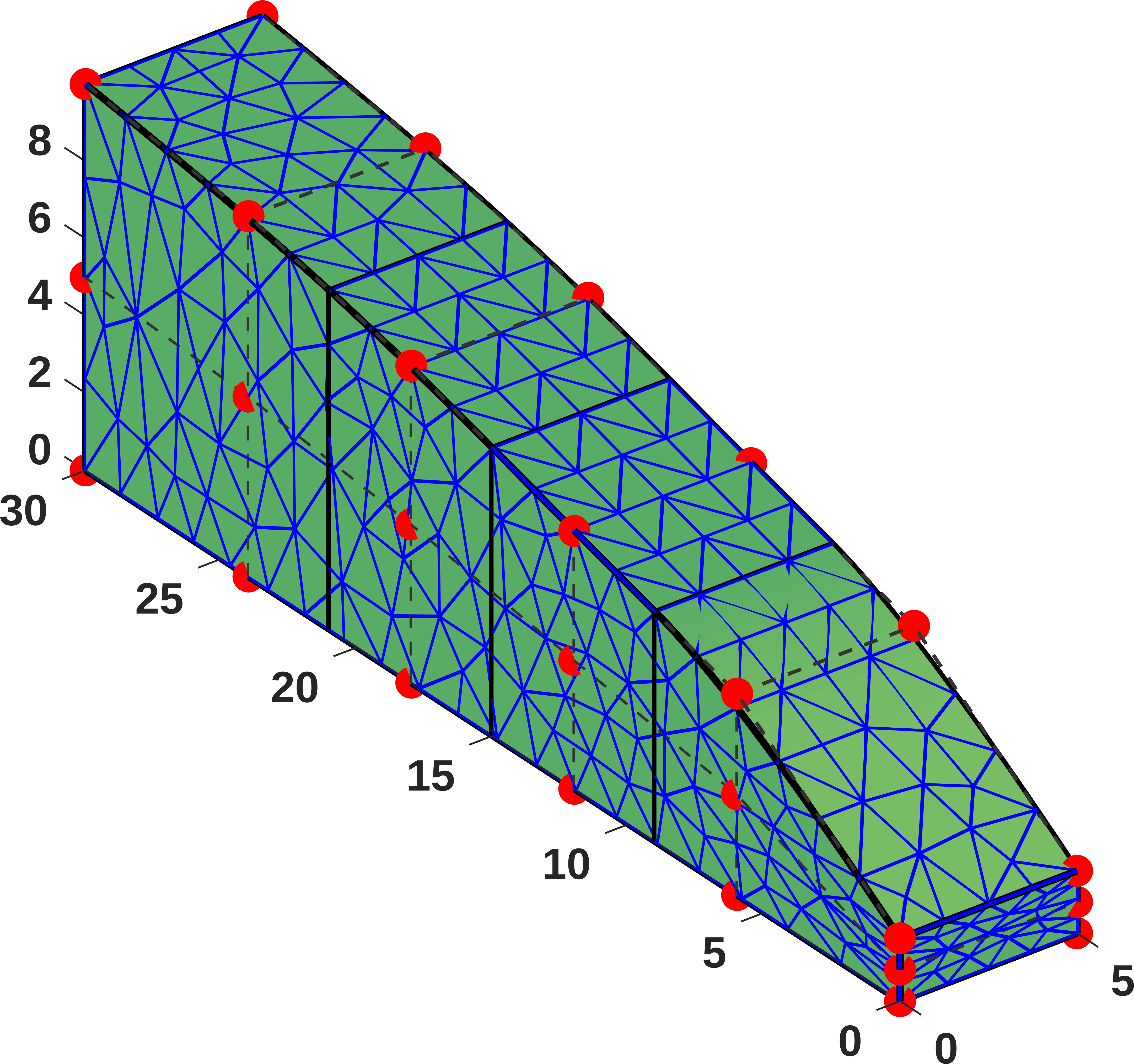}\hfill
   \caption{}
 \end{subfigure}\hspace{15pt}
 \begin{subfigure}[t] {0.4\textwidth}
   \centering
   \includegraphics[width=\textwidth]{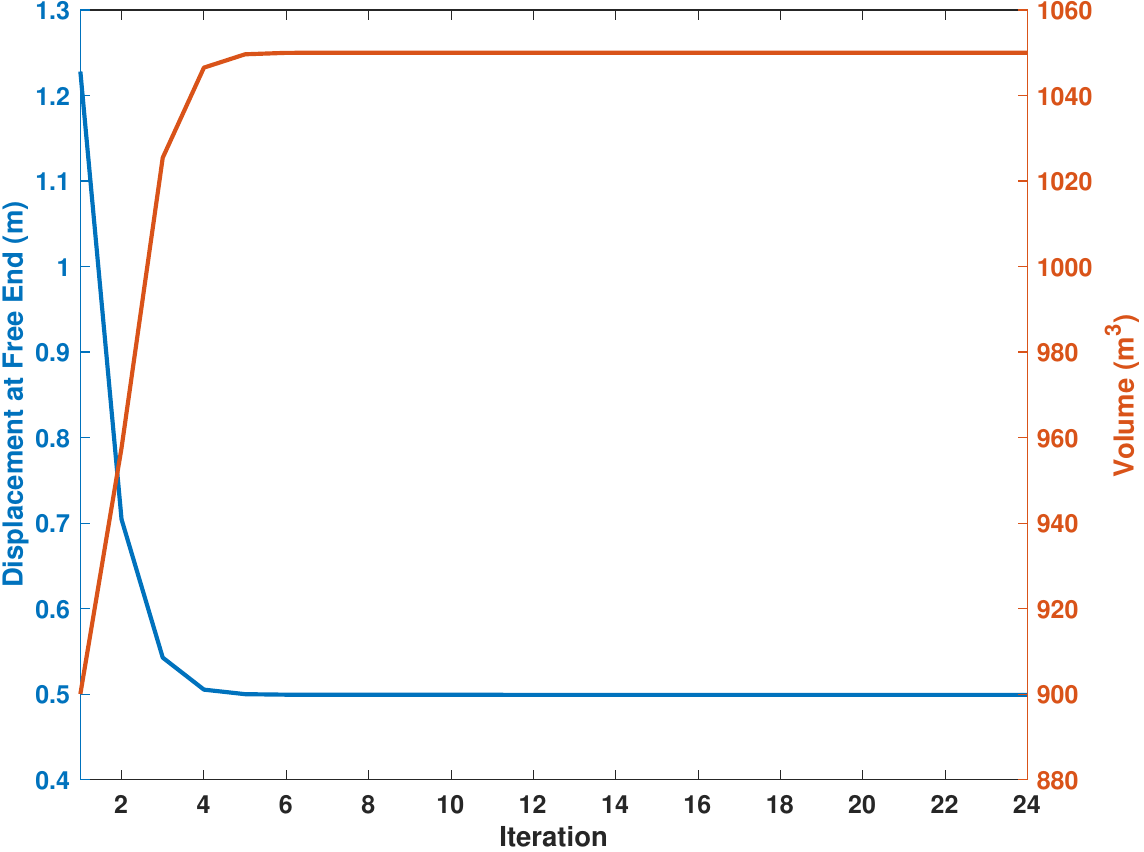}
   \caption{}
 \end{subfigure}
 \caption{Optimal geometries for the cantilever beam problem using cubic B\'{e}zier tetrahedra with size mesh a) $h_l=2$ and c) $h_l=1.5$, and their corresponding convergence curves b), d).}
 \label{fig:CantileverOptimal}
 \end{figure}
 
\subsection{Cantilever beam}
\label{S:7.2}
The first example dealing with shape optimization is a 3D cantilever beam. The initial shape is constructed with 6 tensor product B-spline surfaces, one for each face. The beam has length $L=30$ m, width $b=5$ m and an initial high $h=6$ m. The beam is fixed on one of its faces and a vertical load is applied on the lower edge of the free end, as shown in Fig.~\ref{fig:CantileverInitial}. The problem is formulated as follows:
 
\begin{table}[t]
    \centering
    \caption{Optimization results for the cantilever beam problem.}
    \begin{tabular}{|l|l|l|l|l|l|l|l|l|}
    \hline
    \multicolumn{3}{|c|}{\multirow{3}{*}{\begin{tabular}[c]{@{}c@{}}Numerical\\ experiment\end{tabular}}} & \multicolumn{6}{c|}{Mesh update strategy}                                                                                                                                                                                           \\ \cline{4-9} 
    \multicolumn{3}{|c|}{}                                                                                & \multicolumn{3}{c|}{Remeshing}                                                                                   & \multicolumn{3}{c|}{\begin{tabular}[c]{@{}c@{}}Pseudo-elastic mesh\end{tabular}}  \\ \cline{4-9} 
    \multicolumn{3}{|c|}{}                                                                                & iter                  & $f^{opt}_0$ & \begin{tabular}[c]{@{}l@{}}running\\ time (s)\end{tabular} &iter                  & $f^{opt}_0$ & \begin{tabular}[c]{@{}l@{}}running\\ time (s)\end{tabular} \\ \hline
    \multirow{2}{*}{$p$=2}                       & $h_l$                           & 4                          & \multirow{2}{*}{16} & \multirow{2}{*}{0.4922}     & \multirow{2}{*}{269}                                  & \multirow{2}{*}{23} & \multirow{2}{*}{0.4851}     & \multirow{2}{*}{151}                                \\ \cline{2-3}
                                           & dof's                       & 1203                       &                     &                               &                                                            &                     &                               &                                                            \\ \hline
    \multirow{2}{*}{$p$=2}                       & $h_l$                           & 2                          & \multirow{2}{*}{19} & \multirow{2}{*}{0.4945}     & \multirow{2}{*}{1051}                                  & \multirow{2}{*}{23} & \multirow{2}{*}{0.4918}     & \multirow{2}{*}{596}                                  \\ \cline{2-3}
                                           & dof's                       & 4650                       &                     &                               &                                                            &                     &                               &                                                            \\ \hline
    \multirow{2}{*}{$p$=2}                       & $h_l$                           & 1.5                        & \multirow{2}{*}{19} & \multirow{2}{*}{0.4963}     & \multirow{2}{*}{1595}                                 & \multirow{2}{*}{26} & \multirow{2}{*}{0.4946}     & \multirow{2}{*}{1175}                                  \\ \cline{2-3}
                                           & dof's                       & 7968                       &                     &                               &                                                            &                     &                               &                                                            \\ \hline
    \multirow{2}{*}{$p$=3}                       & $h_l$                           & 4                          & \multirow{2}{*}{20} & \multirow{2}{*}{0.4968}      & \multirow{2}{*}{639}                                  & \multirow{2}{*}{25} & \multirow{2}{*}{0.4951}     & \multirow{2}{*}{252}                                  \\ \cline{2-3}
                                               & dof's                       & 3384                       &                     &                               &                                                            &                     &                               &                                                            \\ \hline
    \multirow{2}{*}{$p$=3}                       & $h_l$                           & 2                          & \multirow{2}{*}{17} & \multirow{2}{*}{0.4988}     & \multirow{2}{*}{2041}                                 & \multirow{2}{*}{24} & \multirow{2}{*}{0.4972}     & \multirow{2}{*}{1050}                                 \\ \cline{2-3}
                                               & dof's                       & 13737                      &                     &                               &                                                            &                     &                               &                                                            \\ \hline
    \multirow{2}{*}{$p$=3}                       & $h_l$                           & 1.5                        & \multirow{2}{*}{17} & \multirow{2}{*}{0.5003}     & \multirow{2}{*}{3143}                                 & \multirow{2}{*}{24} & \multirow{2}{*}{0.4990}     & \multirow{2}{*}{1931}                                 \\ \cline{2-3}
                                           & dof's                       & 23907                      &                     &                               &                                                            &                     &                               &                                                            \\ \hline
    \multirow{2}{*}{$p$=4}                       & $h_l$                           & 4                          & \multirow{2}{*}{18} & \multirow{2}{*}{0.4996}     & \multirow{2}{*}{1092}                                  & \multirow{2}{*}{25} & \multirow{2}{*}{0.4980}     & \multirow{2}{*}{513}                                  \\ \cline{2-3}
                                               & dof's                       & 7275                       &                     &                               &                                                            &                     &                               &                                                            \\ \hline
    \end{tabular}
    \label{table:Cantilever_Results}
\end{table}

\begin{figure}[H]
 \centering
 \begin{subfigure}[t] {0.4\textwidth}
   \centering
   \includegraphics[width=\textwidth]{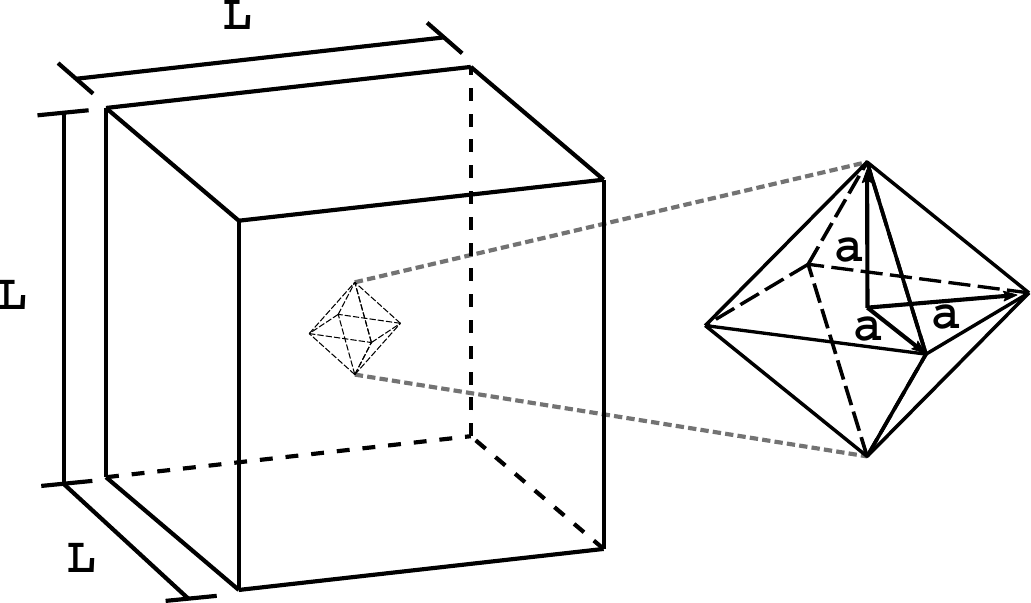}
   \caption{}
   \label{fig:CompleteCubeHoleInitial}
 \end{subfigure}\hspace{15pt}
 \begin{subfigure}[t] {0.25\textwidth}
   \centering
   \includegraphics[width=\textwidth]{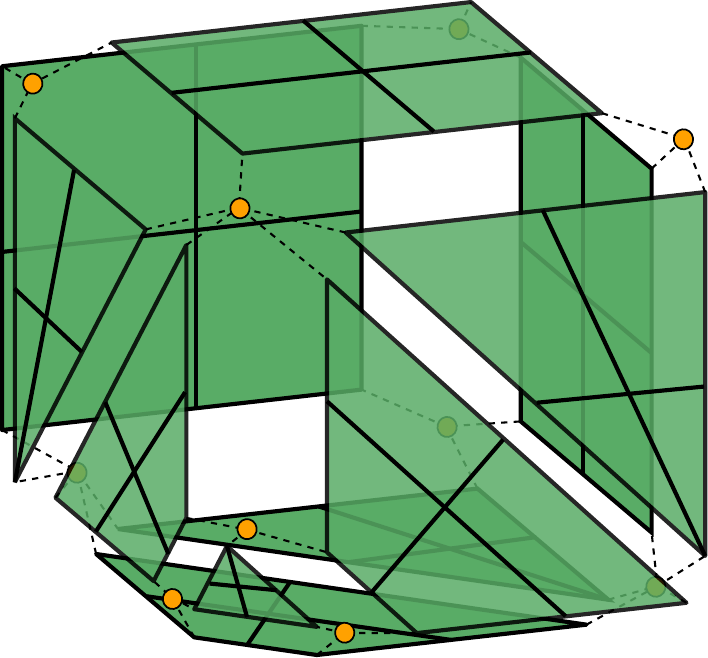}
   \caption{}
 \end{subfigure}\hspace{45pt}
 \begin{subfigure}[t] {0.35\textwidth}
   \centering
   \includegraphics[width=\textwidth]{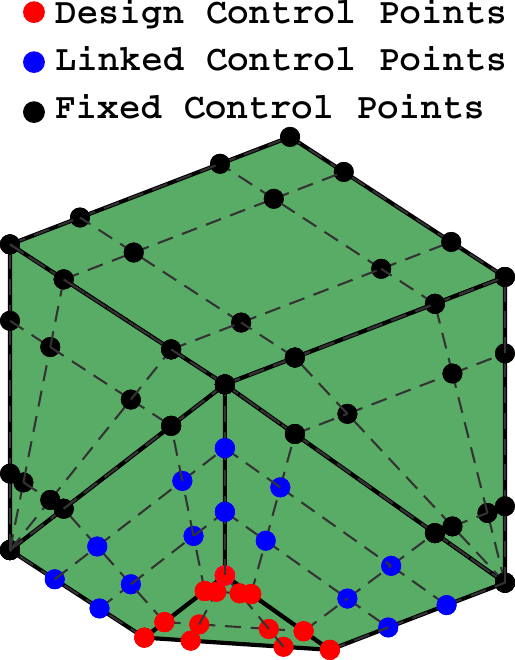}
   \caption{}
   \end{subfigure}\hspace{35pt}
 \begin{subfigure}[t] {0.3\textwidth}
   \centering
   \includegraphics[width=\textwidth]{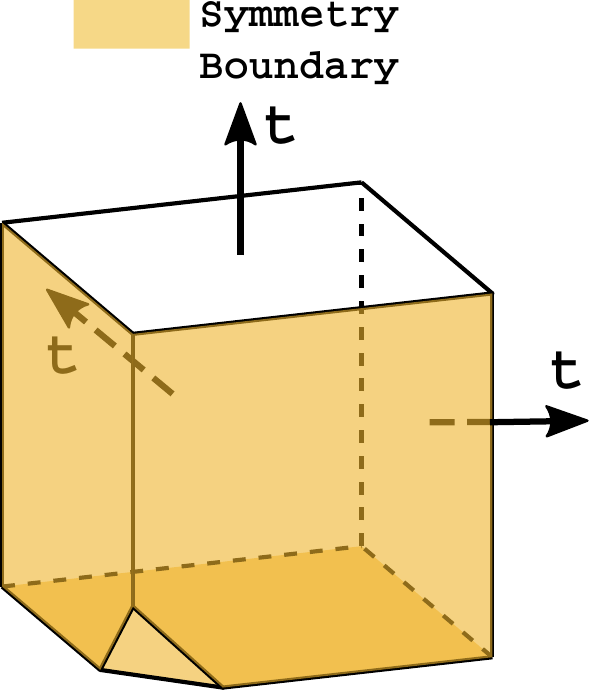}
   \caption{}
   \label{fig:CubeHoleInitialBCs}
 \end{subfigure}
 \caption{a) Dimensions for the cube with a hole example for optimization. b) Initial boundary representation. c) Initial shape for the cube with a hole problem showing the design control points. d) Boundary conditions for an eight of the domain.}
 \label{fig:CubeHoleInitial}
 \end{figure}
 
\begin{equation}
\begin{array}{r@{}l}
&\mbox{minimize} \hspace{45pt}f_0(\bm{s})=\frac{1}{n_{edge}}\sum_{i=1}^{n_{edge}} (u_i^*)^2,\\
&\mbox{such that} \hspace{28pt}V(\bm{s}) \leq \bar{V},
\end{array}
\end{equation}
where $u_i^*$ is the displacement at the $ith$ point over the edge where the load is imposed, $n_{edge}$ refers to the number of control points on the edge and $V(\bm{s})$ is the volume of the domain, which is constrained to a maximum value of $\bar{V} = 1050$ m$^3$. The design variables are the $z$-component of the control points coordinates of the top surface. The design variables are bounded by $1\leq s_i \leq 10$. The MMA parameters used in this example are $\theta=0.1$, $\gamma^{(0)}=0.5$, $\gamma_\ell=1.05$, $\gamma_u=0.65$.

Table~\ref{table:Cantilever_Results} summarizes the results obtained using quadratic, cubic and quartic B\'{e}zier tetrahedra and various mesh sizes. It can be seen that even though the number of iterations required to achieve convergence is lower for the remeshing approach, the use of the mesh update based on the pseudo-elastic mesh reduces the running time considerably, in contrast with remeshing the domain after every iteration has finished. In order to avoid degenerated elements, especially on the free end of the beam, the mesh is updated by calling Gmsh after 8 iterations when the pseudo-elastic approach is used. The number of degrees of freedom shown in Table~\ref{table:Cantilever_Results} refers to the initial mesh.

\begin{figure}[H]
\centering
   \begin{subfigure}[b] {0.3\textwidth}
   \centering
   \includegraphics[width=\textwidth]{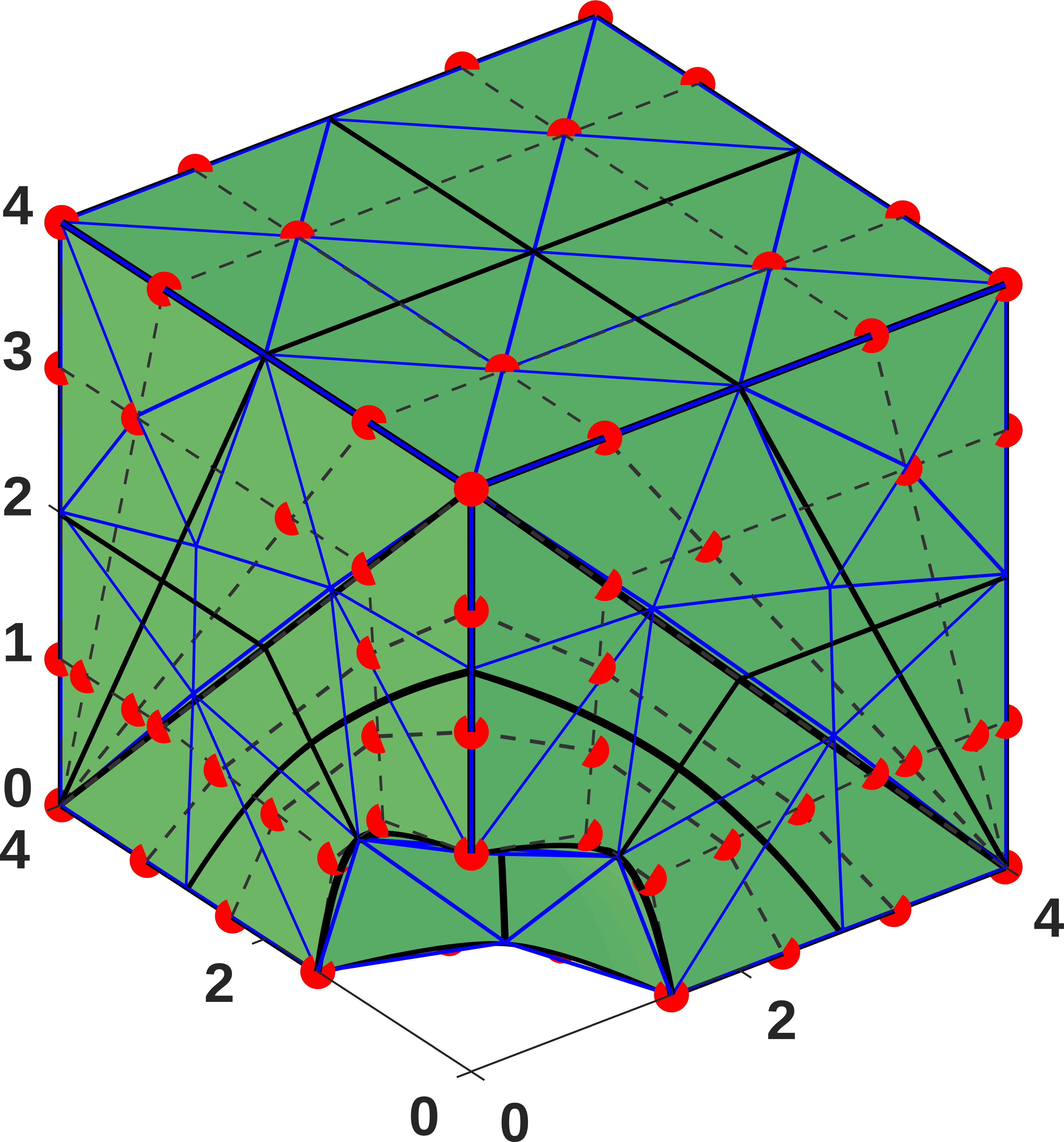}\hfill
   \caption{}
   \label{fig:cubeOptref0}
   \end{subfigure}\hspace{15pt}
   \begin{subfigure}[b] {0.4\textwidth}
   \centering
   \includegraphics[width=\textwidth]{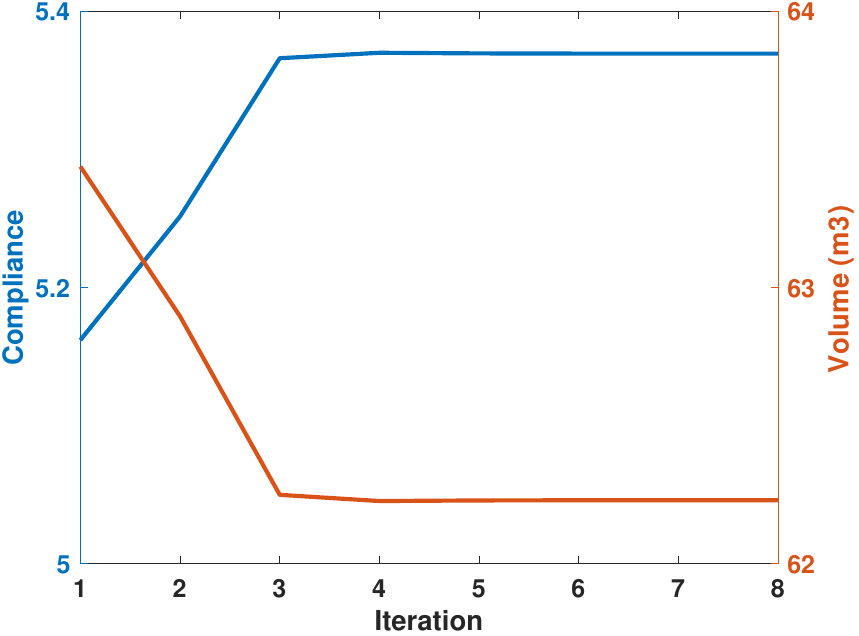}
   \caption{}
   \label{fig:cubeConvergenceref0}
   \end{subfigure}
   
   \begin{subfigure}[b] {0.3\textwidth}
   \centering
   \includegraphics[width=\textwidth]{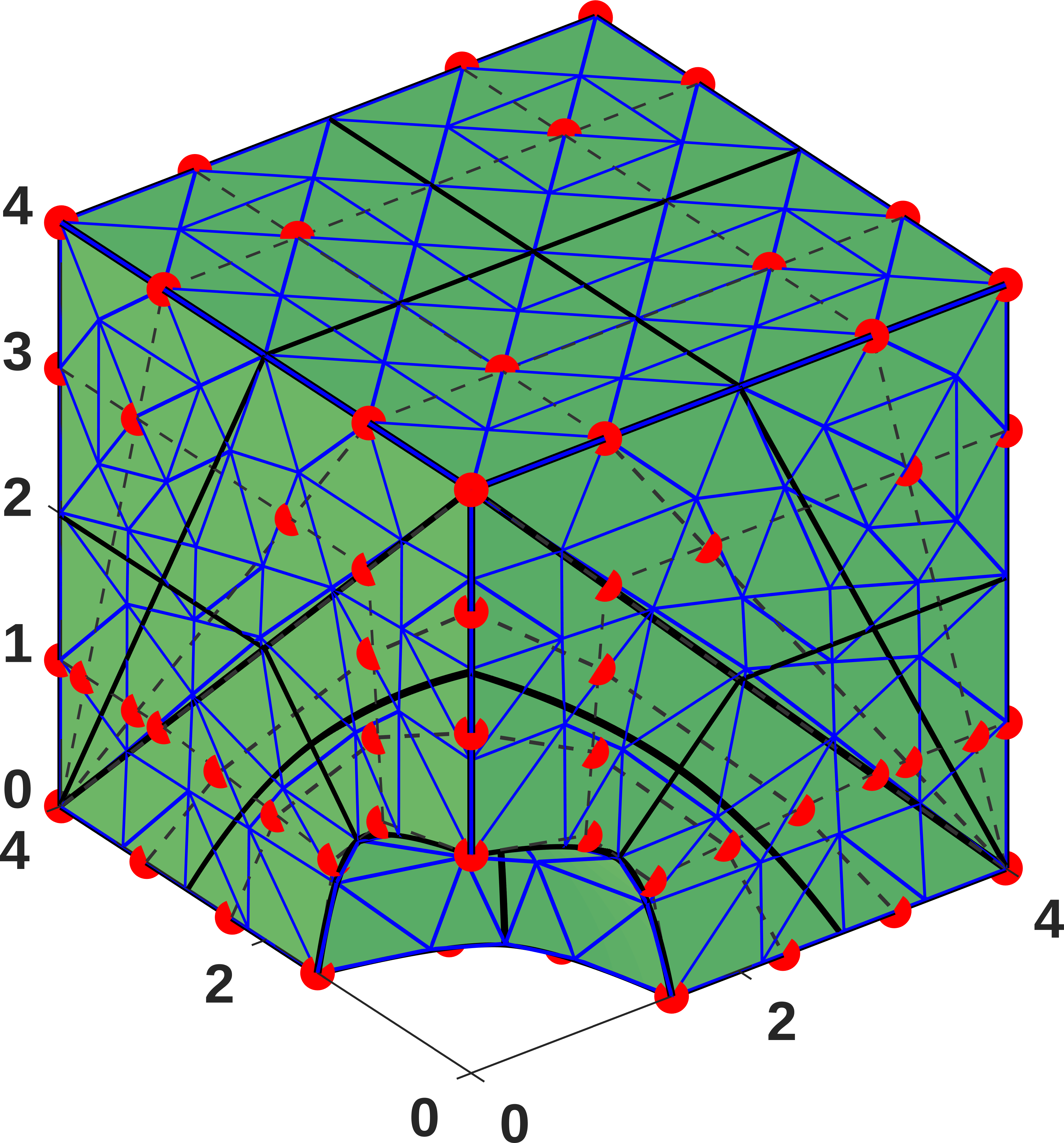}\hfill
   \caption{}
   \label{fig:cubeOptref1}
   \end{subfigure}\hspace{15pt}
   \begin{subfigure}[b] {0.4\textwidth}
   \centering
   \includegraphics[width=\textwidth]{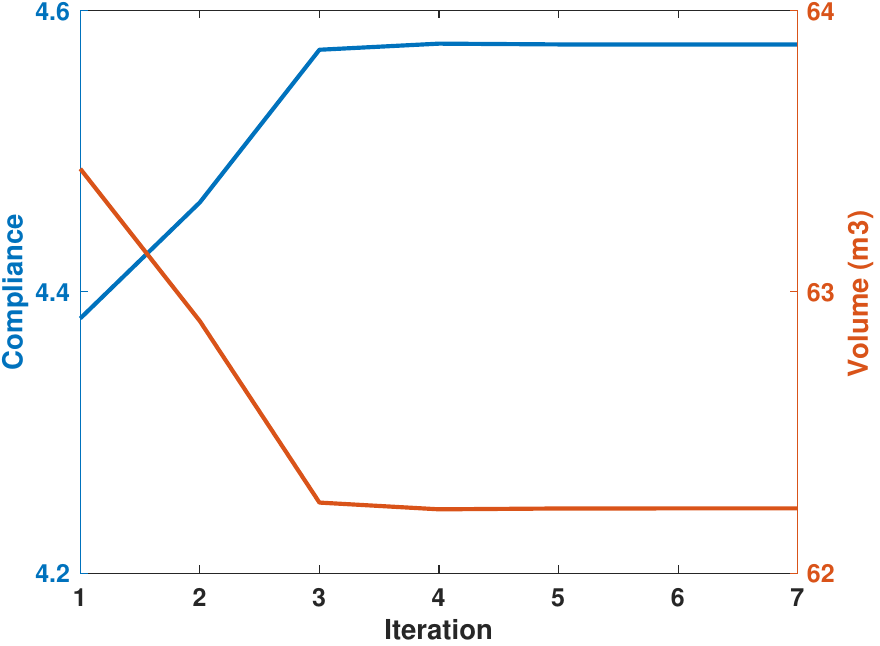}\hspace{15pt}
   \caption{}
   \label{fig:cubeConvergenceref1}
   \end{subfigure}
\caption{Optimal geometries of the cube with a hole example using cubic B\'{e}zier tetrahedra with b) no refinement and c) one refinement, and their corresponding convergence curves b), d).}
\label{fig:CubeHoleInitialSrfs}
\end{figure}

\subsection{Cube with a hole}
\label{S:7.3}
The shape of a hole inside a cube is optimized in order to minimize the strain energy. This example is a 3D version of the plate with a hole commonly used as a benchmark for 2D structural shape optimization. The initial domain consists on a cube with side length $L=4$ with an octahedral inclusion inscribed in a circle of radius $a=1.5$ as shown in Fig.~\ref{fig:CompleteCubeHoleInitial}. Symmetry is considered and therefore one eight of the cube is modelled. A total of 10 surfaces are used for the boundary representation. This example shows that an analysis suitable discretization can be generated even though some of the surfaces include singularities in their parameterization. Fig.~\ref{fig:CubeHoleInitial} show the boundary representation, design variables and boundary conditions of the problem. A traction $t=100$ N is imposed at the outer faces of the cube. The problem consists on minimizing the strain energy with an imposed maximum volume constraint $\bar{V} = 62.23$ m$^3$. The statement is formulated mathematically as:

\begin{equation}
\begin{array}{r@{}l}
&\mbox{minimize} \hspace{30pt}f_0(\bm{s})=\bm{u}^T\bm{f},\\
&\mbox{such that} \hspace{28pt}V(\bm{s}) \leq \bar{V},
\end{array}
\end{equation}

\begin{table}[t]
\centering
\caption{Optimization results for the cube with hole example, where a characteristic length $h_l = 2$ was used over the whole domain.}
\begin{tabular}{|l|l|l|l|l|l|l|l|l|}
\hline
\multicolumn{3}{|c|}{\multirow{3}{*}{Numerical experiment}} & \multicolumn{6}{c|}{Mesh update strategy}                                                                                                                                                                                            \\ \cline{4-9} 
\multicolumn{3}{|c|}{}                                      & \multicolumn{3}{c|}{Remeshing}                                                                                    & \multicolumn{3}{c|}{Pseudo-elastic mesh update}                                                                  \\ \cline{4-9} 
\multicolumn{3}{|c|}{}                                      & iter                  & $f^{opt}_0$ & \begin{tabular}[c]{@{}l@{}}running \\ time (s)\end{tabular} & iter                 & $f^{opt}_0$ & \begin{tabular}[c]{@{}l@{}}running \\ time (s)\end{tabular} \\ \hline
\multirow{2}{*}{p=2}       & refinements       & 0          & \multirow{2}{*}{14} & \multirow{2}{*}{5.3566}       & \multirow{2}{*}{108}                                   & \multirow{2}{*}{7} & \multirow{2}{*}{5.3395}       & \multirow{2}{*}{31}                                    \\ \cline{2-3}
                           & dof's             & 1203       &                     &                               &                                                             &                    &                               &                                                             \\ \hline
\multirow{2}{*}{p=2}       & refinements       & 1          & \multirow{2}{*}{8} & \multirow{2}{*}{4.5746}       & \multirow{2}{*}{293}                                   & \multirow{2}{*}{7} & \multirow{2}{*}{4.5681}       & \multirow{2}{*}{191}                                   \\ \cline{2-3}
                           & dof's             & 4650       &                     &                               &                                                             &                    &                               &                                                             \\ \hline
\multirow{2}{*}{p=3}       & refinements       & 0          & \multirow{2}{*}{11} & \multirow{2}{*}{5.3857}       & \multirow{2}{*}{152}                                   & \multirow{2}{*}{7} & \multirow{2}{*}{5.3702}       & \multirow{2}{*}{53}                                    \\ \cline{2-3}
                           & dof's             & 7968       &                     &                               &                                                             &                    &                               &                                                             \\ \hline
\multirow{2}{*}{p=3}       & refinements       & 1          & \multirow{2}{*}{8} & \multirow{2}{*}{4.5827}       & \multirow{2}{*}{728}                                   & \multirow{2}{*}{7} & \multirow{2}{*}{4.5758}       & \multirow{2}{*}{515}                                   \\ \cline{2-3}
                           & dof's             & 3384       &                     &                               &                                                             &                    &                               &                                                             \\ \hline
\end{tabular}
\label{table:CubeHoleResults}
\end{table}
 
\begin{table}[]
\centering
\caption{Design variables initial values and their corresponding box constraints of the hammer problem.}
\begin{tabular}{|l|l|l|l|}
\hline
\begin{tabular}[c]{@{}l@{}}Design \\ variable $s_i$\end{tabular} & \begin{tabular}[c]{@{}l@{}}Initial \\ value $s^0_i$\end{tabular} & \begin{tabular}[c]{@{}l@{}}Minimum \\ value $s^{min}_i$\end{tabular} & \begin{tabular}[c]{@{}l@{}}Maximum \\ value $s^{max}_i$\end{tabular} \\\hline
$s_1$              & 0                & -2                 & 0.8                \\\hline
$s_2$              & 0                & -2                 & 0.8                \\\hline
$s_3$              & 2                & 1.2                & 3                  \\\hline
$s_4$              & 2                & 1.2                & 3                  \\\hline
$s_5$              & 13               & 11                 & 14                 \\\hline
$s_6$             & 11               & 9                  & 12                 \\\hline
$s_7$              & 11               & 9                  & 12                \\\hline
\end{tabular}
\label{table:HammerDesignVariables}
\end{table}

The design variables are the $x$, $y$ and $z$ coordinates of the control points describing the NURBS surface of the hole with the restrictions imposed by symmetry. The control points at the edges of the hole are restricted to respect symmetric conditions. In total, 27 design variables are considered, with boundary constraints set as $0\leq s_i\leq 1$ for all $s_i$. Cubic elements and one refinement by partitions are used to solve the governing equations. For this example, the MMA parameters were selected as $\theta=0.1$, $\gamma^{(0)}=0.2$, $\gamma_\ell=0.2$, $\gamma_u=0.2$.

The optimal shape approximates to a spherical inclusion as expected, in correspondence with the 2D problem where the optimal shape resembles a circular hole. The results of the optimization are shown in Table~\ref{table:CubeHoleResults}, using quadratic and cubic B\'{e}zier tetrahedra. A constant characteristic length $h_l=2$ was used to control the mesh size. Afterwards, refinement by uniform splitting of the elements was performed. Again, the time required for convergence with a pseudo-elastic mesh is considerably reduced
in comparison to remeshing the domain after every iteration. Moreover, it is shown that the number of iterations is reduced and the objective function is minimized slightly more with the pseudo-elastic approach.

The final shapes using a movable mesh with no refinements and with 1 refinement by splitting are shown in Fig.~\ref{fig:cubeOptref0} and \ref{fig:cubeOptref1} respectively. A comparison of the corresponding convergence curves is shown in Fig.~\ref{fig:cubeConvergenceref0} and Fig.~\ref{fig:cubeConvergenceref1}.

\begin{table}[t]
\centering
\caption{Optimization results for the hammer example.}
\begin{tabular}{|l|l|l|l|l|l|l|l|l|}
\hline
\multicolumn{3}{|c|}{\multirow{3}{*}{Numerical experiment}} & \multicolumn{6}{c|}{Mesh update strategy}                                                                                                                                                                                           \\ \cline{4-9} 
\multicolumn{3}{|c|}{}                                      & \multicolumn{3}{c|}{Remeshing}                                                                                   & \multicolumn{3}{c|}{Pseudo-elastic mesh update}                                                                  \\ \cline{4-9} 
\multicolumn{3}{|c|}{}                                      & it                 & $f^{opt}_0$ & \begin{tabular}[c]{@{}l@{}}running \\ time (s)\end{tabular} & it                 & $f^{opt}_0$ & \begin{tabular}[c]{@{}l@{}}running \\ time (s)\end{tabular} \\ \hline
\multirow{2}{*}{p=2}       & refinements       & 0          & \multirow{2}{*}{8} & \multirow{2}{*}{235.1560}     & \multirow{2}{*}{534}                                     & \multirow{2}{*}{8} & \multirow{2}{*}{233.9033}     & \multirow{2}{*}{156}                                     \\ \cline{2-3}
                           & dof's             & 2571         &                    &                               &                                                             &                    &                               &                                                             \\ \hline
\multirow{2}{*}{p=2}       & refinements       & 1          & \multirow{2}{*}{8} & \multirow{2}{*}{237.9417}     & \multirow{2}{*}{2076}                                    & \multirow{2}{*}{8} & \multirow{2}{*}{237.4860}     & \multirow{2}{*}{950}                                     \\ \cline{2-3}
                           & dof's             & 16107      &                    &                               &                                                             &                    &                               &                                                             \\ \hline
\multirow{2}{*}{p=3}       & refinements       & 0          & \multirow{2}{*}{8} & \multirow{2}{*}{237.8953}     & \multirow{2}{*}{1162}                                    & \multirow{2}{*}{8} & \multirow{2}{*}{237.4585}     & \multirow{2}{*}{348}                                    \\ \cline{2-3}
                           & dof's             & 7389           &                    &                               &                                                             &                    &                               &                                                             \\ \hline
\multirow{2}{*}{p=3}       & refinements       & 1          & \multirow{2}{*}{8} & \multirow{2}{*}{233.7232}     & \multirow{2}{*}{9091}                                     & \multirow{2}{*}{8} & \multirow{2}{*}{238.5702}      & \multirow{2}{*}{5757}                                    \\ \cline{2-3}
                           & dof's             & 49578           &                    &                               &                                                             &                    &                               &                                                             \\ \hline
\end{tabular}
\label{table:HammerResults}
\end{table}

\subsection{Hammer}
\label{S:7.4}

\begin{figure}[p]
 \centering
 \begin{subfigure}[t] {0.4\textwidth}
   \centering
   \includegraphics[width=\textwidth]{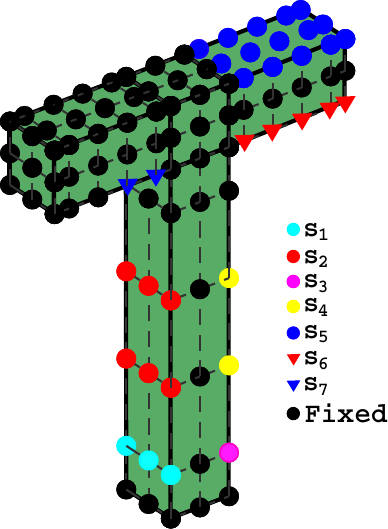}\hfill
   \caption{}
 \end{subfigure}\hspace{15pt}
 \begin{subfigure}[t] {0.5\textwidth}
   \centering
   \includegraphics[width=\textwidth]{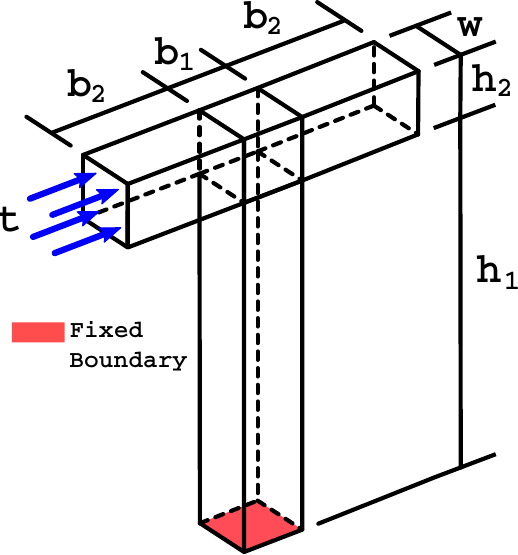}
   \caption{}
   \label{fig:hammerBC}
 \end{subfigure}
 \begin{subfigure}[t] {0.35\textwidth}
   \centering
   \includegraphics[width=\textwidth]{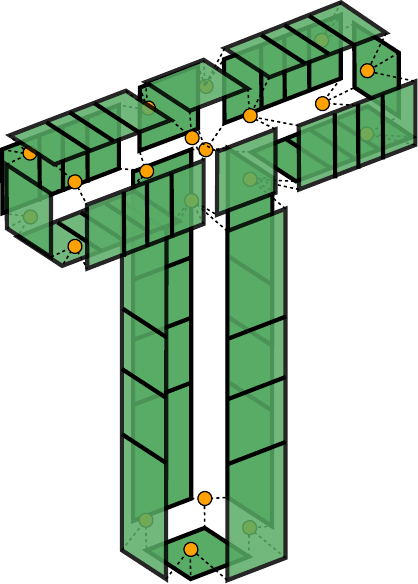}\hfill
   \caption{}
 \end{subfigure}
 \caption{Initial shape of the hammer problem showing a) the design variables, b) boundary conditions and c) set of surfaces used for the boundary representation. }
 \label{fig:HammerInitial}
 \end{figure}

Now, we consider the problem to optimize a hammer, first introduced in \cite{Lian2017}. The problem consists in optimizing a T-shaped geometry to obtain a hammer-like structure by minimizing the strain energy with a volume constraint $V(\bm{s}) \leq 80$ cm$^3$. The design variables, shown on Fig.~\ref{fig:HammerInitial} and Table~\ref{table:HammerDesignVariables}, are the $x$ coordinates of the control points marked as $s_1$, $s_2$, $s_3$ and $s_4$, and the $z$ coordinates for the control points $s_5$, $s_6$ and $s_7$. A traction $t=100$ N is applied on one of the ends of the T-shape, and it is fixed in the bottom face, see Fig.~\ref{fig:hammerBC}.

The initial geometry is constructed with 18 surface patches to describe the boundary. Quadratic NURBS surfaces are used for the boundary representation, and B\'{e}zier tetrahedra for the interior of the domain. The mesh is generated using a homogeneous characteristic length $h_l = 1$. After the mesh is generated, one more refinement by splitting can be performed. Table~\ref{table:HammerResults} shows the optimization results using quadratic and cubic elements with and without refinement. In this case, the same MMA parameters used in the cube with a hole example were also used here.

 \begin{figure}[p]
 \centering
 \begin{subfigure}[b] {0.4\textwidth}
   \centering
   \includegraphics[width=\textwidth]{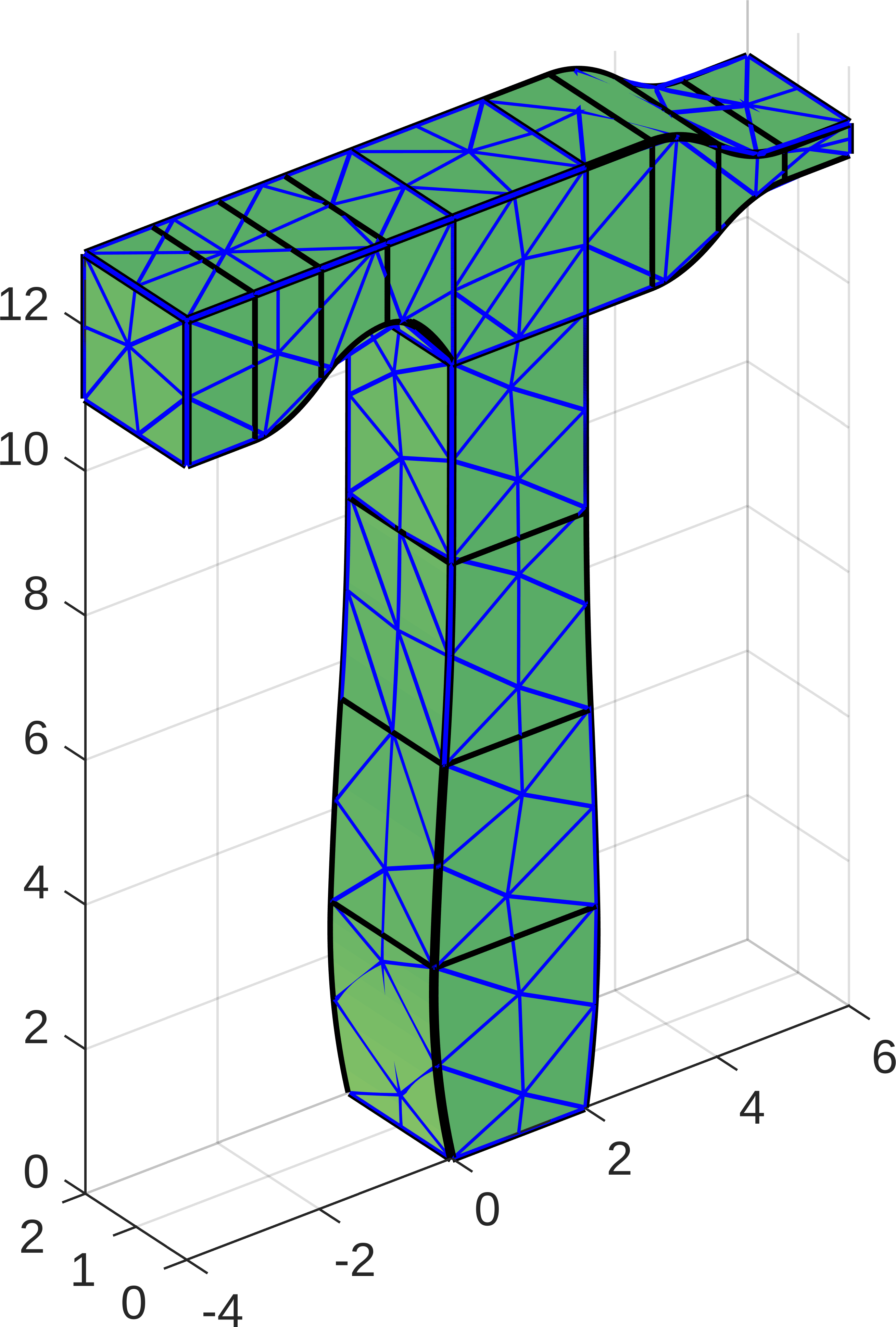}\hfill
   \caption{}
 \end{subfigure}\hspace{15pt}
 \begin{subfigure}[b] {0.5\textwidth}
   \centering
   \includegraphics[width=\textwidth]{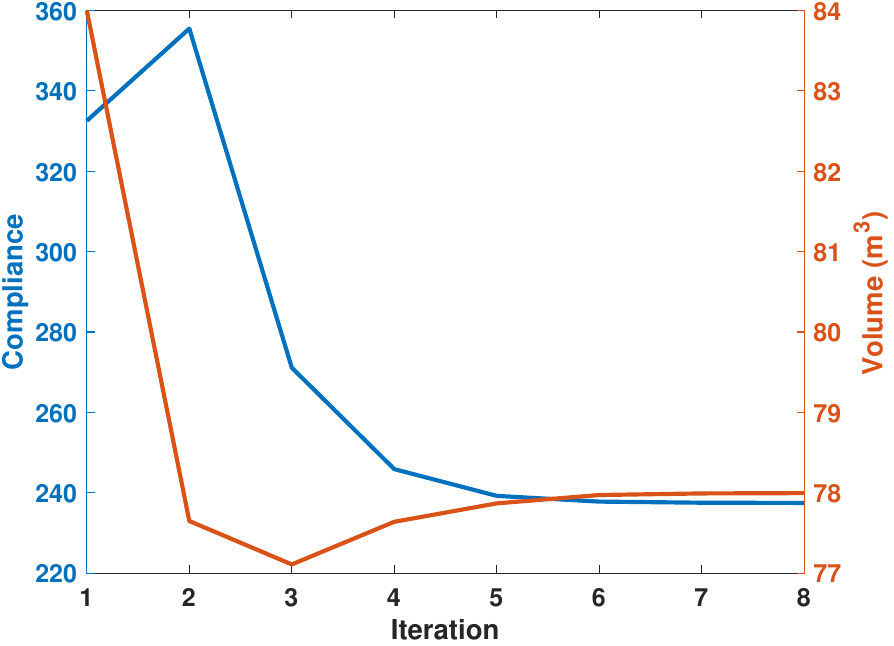}
   \vspace{50pt}
   \caption{}
 \end{subfigure}
 
 \begin{subfigure}[b] {0.4\textwidth}
   \centering
   \includegraphics[width=\textwidth]{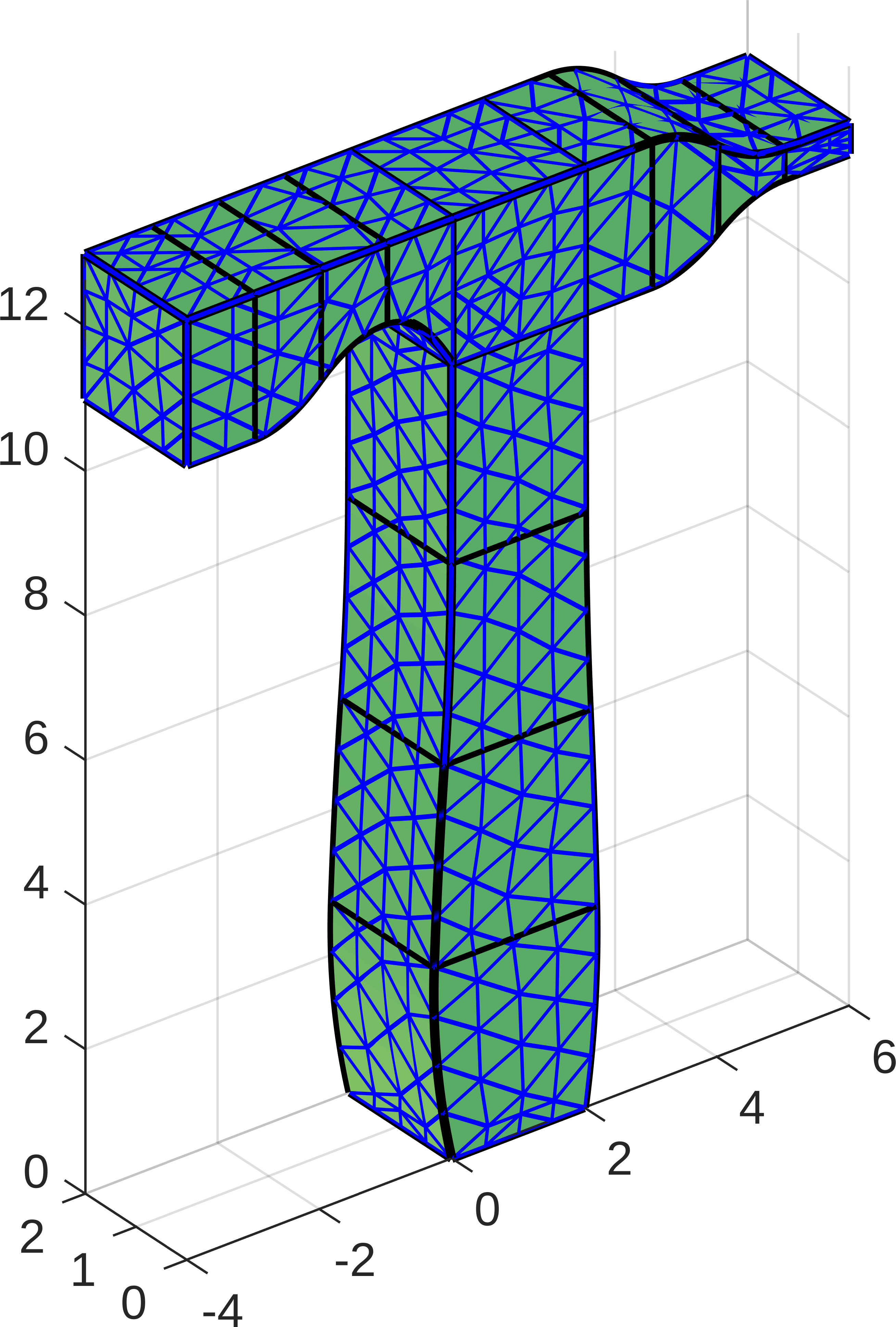}\hfill
   \caption{}
 \end{subfigure}\hspace{15pt}
 \begin{subfigure}[b] {0.5\textwidth}
   \centering
   \includegraphics[width=\textwidth]{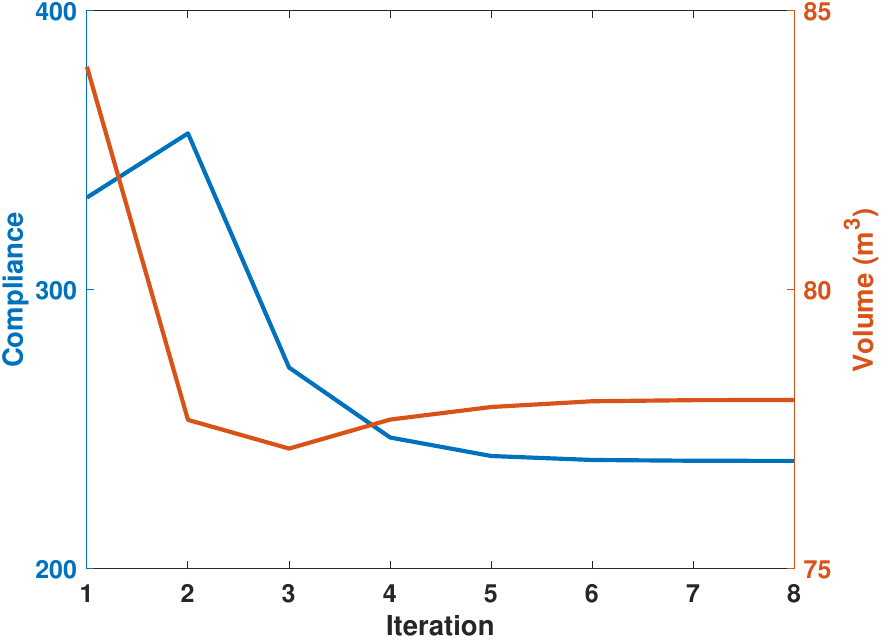}
   \vspace{50pt}
   \caption{}
 \end{subfigure}
 \caption{Optimal shape and convergence results for the hammer example using cubic B\'{e}zier tetrahedra with a), b) No refinement and c), d) 1 refinement.}
 \label{fig:HammerFinal}
 \end{figure}
 
The results in Fig.~\ref{fig:HammerFinal} show that the optimal shape minimizes the strain energy by approximately 29\% of the initial condition. Also the amount of material required is around 92\% of the initial volume. This is in contrast with the results shown in \cite{Lian2017}, where it was reported an increase of the compliance by about 18\% higher than the initial value, constraining the volume constraint to 2\% more than the initial volume. Furthermore, the use of a movable mesh shows again a considerable reduction in the running time to achieve convergence of the optimization problem.

\section{Conclusions}
\label{sec:conclusion}
In this work, we presented a framework for structural shape optimization using CAD-compatible 3D models. NURBS surfaces were used to describe the boundaries of the geometry and B\'{e}zier tetrahedra for the discretization of the domain. The mesh was generated using the software Gmsh, and later the elements were converted into B\'{e}zier elements by computing the coordinates of the surface control points with respect to the initial NURBS description. This framework allows the modeling of complicated geometries using one patch.

An advantage of getting the control points of the B\'{e}zier tetrahedral mesh as a function of the NURBS control points is that the sensitivity analysis for the shape optimization problem, can be performed by implementing analytical sensitivities in the analysis mesh with design variables located in the set of design boundary surfaces.

Additionally, the use of a moving mesh was proposed in order to reuse the mesh originally generated and avoid the use of discretizations with different number of elements along the optimization procedure. This approach results in faster convergence in terms of computational time, though in some cases with a larger number of iterations. Moreover, for the cube with a hole example, the use of the pseudo-elastic mesh update approach also achieves convergence with a reduced number of iterations and minimizing slightly more the objective function.

Possible extensions of this work include the use of $C^1$ continuous elements and its implementation for other type of physical problems. The use of stress constraints can also be considered. Moreover, improvements in the boundary triangulation approach as well as for the mesh update can be done in order to avoid remeshing for longer periods.

\section{Acknowledgements}
Jorge L\'{o}pez appretiates the support of the Country-related cooperation programme with Mexico: CONACYT-DAAD funding programme number 57177537. The authors acknowledge Prof. Svanberg from KTH Royal Institute of Technology, Stockholm, for providing the MMA toolbox. The research leading to these results has received funding from the Deutsche Forschungsgemeinschaft (DFG) - Project number 392023639.

\bibliographystyle{abbrvnatnourl}
\bibliography{sample}

\end{document}